\documentclass[%
 twocolumn,%
 aip,
 jcp,
 amsmath,amssymb,
 reprint,%
 groupedaddress
]{revtex4-2}

\usepackage{graphicx}%
\usepackage{dcolumn}%
\usepackage{bm}%
\usepackage{color}
\usepackage{float}
\usepackage{epsfig}
\usepackage{lipsum}
\usepackage{CJKutf8} %

\usepackage{array}
\usepackage{multirow}
\usepackage[
  colorlinks=true,
  linkcolor=red,    %
  citecolor=blue,   %
  urlcolor=blue    %
]{hyperref}

\newcolumntype{Y}{>{\centering\arraybackslash}X}

\newcommand{\Sec}[1]{Sec.~\ref{#1}}

\newcommand{\be}{\begin{equation}}
\newcommand{\ee}{\end{equation}}
\newcommand{\bea}{\begin{eqnarray}}
\newcommand{\eea}{\end{eqnarray}}
\newcommand{\Fig}[1]{Fig.~\ref{#1}}
\newcommand{\Tab}[1]{Table~\ref{#1}}
\newcommand{\Eq}[1]{Eq.~\eqref{#1}}

\newcommand{\rcite}[1]{Ref.~\onlinecite{#1}}

\newcommand{\B}[1]{\ensuremath{\mathbf{#1}}}

\newcommand{\irrep}[1]{\ensuremath{\mathrm{#1}}}

\newcommand{\symop}[1]{\ensuremath{\mathrm{#1}}}

\begin{document}
\begin{CJK*}{UTF8}{gbsn}
\preprint{AIP/123-QED}

\title{Exact tunneling splittings from path-integral hybrid Monte Carlo with enveloping bridging potentials}

\author{Yu-Chen Wang (汪宇晨)}
\email{wangyuc@phys.chem.ethz.ch}

\author{Jeremy O. Richardson}
\affiliation{\mbox{Institute of Molecular Physical Science, ETH Z\"{u}rich, 8093 Z\"{u}rich, Switzerland}}

\begin{abstract}
A path-integral hybrid Monte Carlo approach with enveloping bridging potentials (PIHMC-EBP) is proposed for calculating numerically exact tunneling splittings in molecular systems.
The central idea is to construct an approximately barrierless bridging potential that smoothly connects symmetry-related regions of ring-polymer phase space, enabling direct sampling of the free-energy profile from which the relevant splittings are obtained.
Two tailored nonlocal updates are designed to enhance the sampling of slow collective motions.
Compared with path-integral molecular dynamics using thermodynamic integration, PIHMC-EBP obviates the need for quadrature and time-step convergence checks, thereby substantially reducing the manual effort required to analyze the results.
Applications to malonaldehyde (and its deuterated isotopologue) and the HCl dimer using state-of-the-art potential energy surfaces provide the most precise tunneling splittings reported to date for both systems, while simultaneously reducing the overall computational cost by several times and three orders of magnitude, respectively.
Finally, application to the water dimer yields the first numerically exact path-integral calculations of the ground-state tunneling splittings on three different potential energy surfaces, all obtained simultaneously by reweighting a single set of trajectories.
\end{abstract}

\maketitle
\end{CJK*}

\section{INTRODUCTION}
Quantum tunneling between symmetry-related molecular configurations breaks the degeneracy of rovibrational levels and produces characteristic tunneling splittings.\cite{Benderskiui-ChemicalDynamicsLow-1994,Bell-TunnelEffectChemistry-1980,Bunker-MolecularSymmetrySpectroscopy-2006,Bunker-FundamentalsMolecularSymmetry-2005}
These splittings act as spectral fingerprints that are exquisitely sensitive to the shape of the potential energy surface (PES) along tunneling pathways through barrier regions, thus providing a powerful probe of large-amplitude motion and PES features that are otherwise difficult to access experimentally.
In modern high-resolution spectroscopy, accurate theoretical calculations are indispensable in predicting, assigning, and quantitatively interpreting tunneling splittings.\cite{Prager-CR-1997-2933,Keutsch-PNASU-2001-10533,Richardson-S-2016-1310,Schwan-ACIE-2019-13119}

Theoretical computations of tunneling splittings typically require evaluating the PES at many geometries.
Since electronic-structure calculations with spectroscopic accuracy are very expensive, a widely used strategy is to fit a high-quality analytic PES from benchmark \textit{ab initio} data and then compute tunneling splittings (or more generally rovibrational energy levels) on that surface.\cite{Wang-JCP-2008-224314,Mizukami-JCP-2014-144310,Qu-ARPC-2018-151,Bowman-JCP-2025-180901,Jing-JPCL-2025-10923,Shen-JCP-2025-154307}
The resulting splittings serve as an important and stringent benchmark of the accuracy of the PES\@.
For instance, vibration-rotation-tunneling splittings have long been used to benchmark and refine the intermolecular potential of water.\cite{Fellers-S-1999-945,Goldman-JCP-2002-10148,Huang-JPCA-2006-445,Huang-JCP-2008-34312,Shank-JCP-2009-144314,Mukhopadhyay-CPL-2018-163,Jing-JPCL-2025-10923,Zhu-JCTC-2023-3551,Babin-JCTC-2013-5395,Leforestier-JCP-2012-14305}
Therefore, high-accuracy splitting calculations not only complement experiment but also provide sensitive feedback for advancing electronic-structure methods and PESs.\cite{Stone-TheoryIntermolecularForces-2013}

One direct way to calculate tunneling splittings is to solve the time-independent nuclear Schr\"{o}dinger equation variationally.
For systems beyond triatomics, the required numerical resource grows exponentially with system size, necessitating numerical schemes like Smolyak sparse grids and matrix-free iterative eigensolvers.\cite{Wang-JCP-2018-74108,Carrington-JCP-2017-120902,Wang-JCP-2008-234102,Sunaga-JCTC-2024-8100,Lauvergnat-C-2023-202300501,Felker-JCP-2023-234109,Simko-JCP-2025-34301}
A complementary route is provided by time-dependent approaches, which extract energy levels from imaginary- or real-time propagators.\cite{Hoppe-JCP-2024-34104,Hammer-JCP-2012-54105,Hammer-JCP-2011-224305,Schroeder-JCP-2011-234307}
These wavefunction methods have delivered highly accurate reference splittings for polyatomic molecules and clusters where fully converged calculations are feasible, but their cost still rises steeply with dimensionality.
To avoid explicit wavefunction construction, diffusion Monte Carlo (DMC) projects out ground and excited states in imaginary time using random walkers, with excited states accessed by imposing a nodal surface.\cite{Gregory-JCP-1995-7817,Quack-CPL-1995-71,Coutinho-Neto-JCP-2004-9207,Wang-JCP-2008-224314,Mizukami-JCP-2014-144310}
However, the results can be subject to systematic bias since the nodal constraint is usually approximate, and the splittings may be poorly resolved statistically when they are very small.

The path-integral formalism\cite{Feynman-QuantumMechanicsPath-1965} provides an alternative route that bypasses explicit wavefunction construction.
Applying a steepest-descent approximation to this integral yields the semiclassical instanton approximation\cite{Benderskiui-ChemicalDynamicsLow-1994,Mil’nikov-JCP-2001-6881}, whose discretized form is known as the ring-polymer instanton (RPI) approach.\cite{Richardson-JCP-2011-54109,Richardson-JCP-2011-124109,Richardson-S-2016-1310,Richardson-IRPC-2018-171,Richardson-PCCP-2017-966,Sahu-JCC-2021-210,Kaeser-JCTC-2022-6840,Nandi-JACS-2023-9655,Sahu-JCC-2021-210,Videla-JPCL-2023-6368,Kaeser-JCTC-2025-6633,Zhang-JCTC-2025-7517,Fiechter-JCP-2026-124115}
In this method, the path integral is formulated in terms of ring-polymer configurations, i.e., closed chains of replicas (“beads”) connected by harmonic springs.
Despite its favorable cost--accuracy balance, instanton theory relies on a harmonic treatment of fluctuations transverse to the tunneling path, which leads to significant errors when anharmonicity is strong, although perturbative corrections\cite{Lawrence-JCP-2023-14111} offer systematic improvements when the deviations are moderate.

Two approaches that go beyond the limitations of the aforementioned methods are path-integral Monte Carlo (PIMC) and path-integral molecular dynamics (PIMD).
Both approaches are, in principle, numerically exact within statistical uncertainty, yet typically more expensive than instanton theory because they sample the full ring-polymer configuration space.
With carefully designed global updates, PIMC has been widely applied to lattice models and quantum many-body systems\cite{Chandler-JCP-1981-4078,Ceperley-RMP-1995-279} (\textit{e.g.}, bosonic fluids).
By contrast, PIMD introduces fictitious momenta for the beads to enable molecular-dynamics sampling in an extended phase space\cite{Parrinello-JCP-1984-860,Berne-ARPC-1986-401,Marx-JCP-1996-4077,Tuckerman-StatisticalMechanicsTheory-2010}, and is therefore well suited to molecular systems.
M\'{a}tyus and co-workers\cite{Matyus-JCP-2016-114108,Matyus-JCP-2016-114109} used PIMD to calculate ratios of thermal density-matrix elements connecting symmetry-related minima, and extracted the splittings by fitting the temperature dependence of the ratios at sufficiently low temperatures, where only the ground and first excited states contribute appreciably.
This approach scales favorably in computational cost with system size and treats anharmonicity exactly. %
It has been applied to compute the tunneling splitting of malonaldehyde\cite{Matyus-JCP-2016-114108,Vaillant-JCP-2018-234102} and a few water clusters.\cite{Vaillant-JCP-2018-234102,Vaillant-JPCL-2019-7300,Zhu-JACS-2022-21356,Zhu-JCP-2023-220901}
However, small rotational level spacings can lead to significant contamination from rotationally excited states,  often making the results unreliable.

Recently, our group\cite{Trenins-JCP-2023-34108} proposed replacing the density-matrix elements with symmetrized partition functions and introduced an ``Eckart spring'' into the ring-polymer potential.
This modification allows the splitting to be obtained from the calculation at a single temperature and rigorously projects the system onto the rotational ground state, thereby eliminating the contamination problem and appreciably improving the efficiency.
Subsequent application\cite{Baumann-MP-2025-2474202} to malonaldehyde delivered a statistically converged tunneling splitting with a substantially smaller uncertainty than previous wavefunction-based and DMC benchmarks.
Very recently, this scheme has been extended to project onto specific rotational states and obtain rotationally excited tunneling splittings.\cite{Zupan-ExactTunnelingSplittings-2026}

Different symmetrized partition functions (and, similarly, density-matrix elements) correspond to different effective ring-polymer potentials.
As a result, the associated ensembles are weighted by their respective Boltzmann factors. 
This implies that the ratio of two such quantities can be obtained from the free-energy difference between their effective potentials.
In a straightforward PIMD simulation, only a single potential is sampled, leaving the free-energy difference inaccessible unless the relevant ensembles exhibit substantial overlap.
Consequently, prior studies\cite{Matyus-JCP-2016-114108,Matyus-JCP-2016-114109,Vaillant-JCP-2018-234102,Vaillant-JPCL-2019-7300,Zhu-JACS-2022-21356,Zhu-JCP-2023-220901,Trenins-JCP-2023-34108,Baumann-MP-2025-2474202,Zupan-ExactTunnelingSplittings-2026} have combined PIMD with thermodynamic integration (TI), wherein a thermodynamic path is chosen to continuously transform one potential into the other.
In practice, TI is discretized into quadrature points, and a separate PIMD simulation is run at each point. %
The results are later combined to obtain the free-energy difference.

Despite the previous successes of PIMD-TI, this workflow introduces several numerical complications.
First, the discretization error of the quadrature is difficult to predict \textit{a priori}, often necessitating repeated calculations with increasing numbers of quadrature points to ensure convergence.
Second, near the endpoints of the path, the relevant estimators often have large variances with opposite signs; when integrated, their partial cancellation inflates the statistical uncertainty, although we note that carefully designed integration paths may mitigate this effect.\cite{Srdinsek-PRR-2022-32002}
Third, in multiwell systems with large symmetry groups, such as water dimer, multiple TIs must be evaluated, and each inherits the same difficulties, further increasing complexity.\cite{Vaillant-JCP-2018-234102}
Finally, PIMD integrators employ a finite time step, $\Delta t$, which introduces a systematic bias and typically requires additional simulations at smaller $\Delta t$ to demonstrate that this bias is negligible.
Taken together, these factors make a full PIMD-TI calculation very demanding not only in computational cost but also in terms of human effort.
As such, further improvements are required to overcome the above difficulties and enable an extensive study of tunneling for a wide range of molecules.

In this paper, we introduce a new path-integral approach for the calculation of tunneling splittings, termed path-integral hybrid Monte Carlo with enveloping bridging potentials (PIHMC-EBP).
The method replaces the TI used in earlier PIMD studies by constructing an effective potential that allows the relevant free-energy profile to be obtained more directly.
Combined with HMC sampling and tailored nonlocal updates, this yields a simpler and more efficient computational procedure, as demonstrated below.

The remainder of this paper is organized as follows:
\Sec{sec2} presents the theoretical foundation of PIHMC-EBP for calculating tunneling splittings;
\Sec{sec3} demonstrates the performance of this method through applications to malonaldehyde, the HCl dimer, and the water dimer, with comparisons to existing benchmarks;
\Sec{sec4} provides further discussion and concluding remarks.

\section{Methodology\label{sec2}}
This section summarizes the theoretical framework and computational strategy used in the present work.
Sections~\ref{sec:zpj} and \ref{sec:rp} review the symmetrized path-integral framework with Eckart springs established in earlier studies\cite{Trenins-JCP-2023-34108,Zupan-ExactTunnelingSplittings-2026} and its connection to tunneling splittings.
Section~\ref{sec:ebp} introduces the details of the EBP formulation.
Section~\ref{sec:pihmc} describes the PIHMC sampling scheme, while \Sec{sec:post} discusses several technical aspects of the implementation.
Finally, \Sec{sec:workflow} outlines the overall workflow of the method. Readers already familiar with the symmetrized path-integral framework may proceed directly to \Sec{sec:ebp}.
\subsection{Symmetrized partition function}\label{sec:zpj}
\subsubsection{Projection operator}
Consider an isolated molecular system described by a rovibrational Hamiltonian $\hat{H}$ that is invariant under a set of discrete symmetry operations.
These operations form a finite group $\mathcal{G}$ of order $|\mathcal{G}|$, referred to as the molecular symmetry group of the Hamiltonian.\cite{Bunker-FundamentalsMolecularSymmetry-2005,Bunker-MolecularSymmetrySpectroscopy-2006}
Each symmetry operation $\hat{P}\in\mathcal{G}$ is represented by a unitary operator that commutes with $\hat{H}$, so that the eigenstates of $\hat{H}$ can be chosen to transform according to irreducible representations (irreps) of $\mathcal{G}$.
Symmetry operations in molecular systems include permutations of indistinguishable nuclei, inversion with respect to the center of mass, and their combinations. %

We denote the eigenstates of $\hat{H}$ by $\lvert n,\ell\rangle$, where $n$ labels the energy level with eigenvalue $E_n$, and $\ell$ distinguishes degenerate states within that level.
For a given $n$ and in the absence of accidental degeneracies, the set of states $\{\lvert n,\ell\rangle\}$ spans an invariant subspace under the action of all $\hat{P}\in\mathcal{G}$.
These states transform according to an irrep $\Gamma$ of dimension $d_\Gamma$, and they may be labeled as $\ell=1,\dots,d_\Gamma$.
In this way, the structure of the rovibrational spectrum reflects the irreps of $\mathcal{G}$, and tunneling splittings between symmetry-adapted levels can be expressed compactly in group-theoretical terms.
A central tool in this construction is the projection operator onto a specific irrep $\Gamma$ given by\cite{Bunker-MolecularSymmetrySpectroscopy-2006,Bunker-FundamentalsMolecularSymmetry-2005}
\begin{equation}
\label{eq:proj-gamma}
\mathcal{P}^{(\Gamma)}
= \frac{d_\Gamma}{|\mathcal{G}|}
\sum_{\hat{P}\in\mathcal{G}} \chi^{(\Gamma)*}_P\,\hat{P},
\end{equation}
where $\chi^{(\Gamma)}_P$ is the character of $\hat{P}$ in $\Gamma$.
The operator $\mathcal{P}^{(\Gamma)}$ is Hermitian and idempotent, projects onto all states transforming as $\Gamma$, and commutes with $\hat{H}$.

Apart from discrete symmetry operations, the molecular Hamiltonian is also invariant under spatial rotation in the absence of external fields.
As such, the eigenfunctions of the Hamiltonian can be labeled by the total angular momentum quantum number $J$ and its space-fixed projection $M=-J,\dots,J$.\footnote{We assume a closed-shell molecular system, for which interactions involving electronic spins vanish. We also neglect hyperfine interactions from nuclear spins.}
For each value of $J$, the $(2J+1)$ states are exactly degenerate and form an irrep of the rotation group, adding an additional layer of degeneracy beyond that from the discrete symmetry.
The corresponding projection operator onto the subspace with total angular momentum $J$ is\cite{Ring-NuclearManyBody-2004,Zare-UnderstandingSpatialAspects-1988}
\begin{equation}
\label{eq:PJ-def}
\mathcal{P}^{(J)}
= \frac{2J+1}{8\pi^2}
\int \mathrm{d}\mathbf{\Omega}\,
\chi^{(J)}(\mathbf{\Omega})\,
\hat{R}(\mathbf{\Omega}),
\end{equation}
where $\hat{R}(\mathbf{\Omega})$ denotes an overall rotation of the system by the angles $\mathbf{\Omega}$ in a space-fixed frame with origin at the center of mass\footnote{Here $\mathbf{\Omega}$ represents a parametrization of the three-dimensional rotation group. The integration over $\mathbf{\Omega}$ may be written in terms of any standard set of rotational angles, such as Euler angles or Tait-Bryan angles, provided the appropriate integration measure is used.}, and $\chi^{(J)}(\mathbf{\Omega})$ is the character (trace) of the Wigner $D$-matrix for angular momentum $J$.\footnote{The trace of the Wigner $D$-matrix is always real.}
This character can be  calculated according to
\begin{equation}
\chi^{(J)}(\mathbf{\Omega})
= \frac{\sin\!\left[(J+\tfrac{1}{2})\theta\right]}
       {\sin\!\left(\tfrac{\theta}{2}\right)},
\end{equation}
and the angle $\theta\in[0,\pi]$ is obtained directly from $\hat{R}(\mathbf{\Omega})$ via
\begin{equation}
\cos\theta=\frac{1}{2}\left(\mathrm{Tr}[\hat{R}(\mathbf{\Omega})]-1\right).
\end{equation}
The operator $\mathcal{P}^{(J)}$ projects onto all states with total angular momentum $J$, independently of $M$, and commutes with both $\hat{H}$ and the elements of $\mathcal{G}$.

\subsubsection{Rotationally-Projected symmetrized partition function}
We now combine the discrete-symmetry projector $\mathcal{P}^{(\Gamma)}$ and the rotational projector  $\mathcal{P}^{(J)}$ with the Boltzmann operator $e^{-\beta \hat{H}}$ (\textit{i.e.} the unnormalized thermal density operator).
Taking the trace yields
\begin{equation}
\label{eq:proj-trace}
\mathrm{Tr}\!\left[e^{-\beta\hat{H}}\mathcal{P}^{(\Gamma)} \mathcal{P}^{(J)}\right]
= d_\Gamma (2J+1)\sum_{n\in\Gamma,J} e^{-\beta E_n},
\end{equation}
where $\beta = 1/(k_{\mathrm{B}}T)$ and the sum runs over all distinct energy levels $n$ with total angular momentum $J$ and symmetry $\Gamma$.
Thus, the combined projection isolates the $(\Gamma,J)$-block of the spectrum and counts each level with a factor $d_\Gamma(2J+1)$.

In the low-temperature limit,  \Eq{eq:proj-trace} gives direct access to tunneling splittings between irreps.
Let $E^{(\Gamma,J)}$ and $E^{(\Gamma',J')}$ be the lowest energies belonging to irreps $\Gamma$ and $\Gamma'$ with total angular momentum $J$ and $J'$, respectively, and define the corresponding tunneling splitting $\Delta_{\Gamma\Gamma'}^{JJ'}=E^{(\Gamma,J)} - E^{(\Gamma',J')}$.
From \Eq{eq:proj-trace} one obtains
\begin{equation}
\label{eq:delta-proj}
e^{-\beta\Delta_{\Gamma\Gamma'}^{JJ'}}
= \frac{\mathrm{Tr}\!\left[e^{-\beta\hat{H}}\mathcal{P}^{(\Gamma)}\mathcal{P}^{(J)}\right]/[d_\Gamma(2J+1)]}
       {\mathrm{Tr}\!\left[e^{-\beta\hat{H}}\mathcal{P}^{(\Gamma')}\mathcal{P}^{(J')}\right]/[d_{\Gamma'}(2J'+1)]}
\end{equation}
for $\beta\to\infty$.

To connect this formulation to quantities accessible in path-integral simulations, it is useful to introduce $J$-projected symmetrized partition functions associated with individual symmetry operations.
For any $\hat{P}\in\mathcal{G}$, define\footnote{Summing over all $J$ recovers the unprojected symmetrized partition function, $\sum_J(2J+1)Z_P^{(J)}=Z_P$, defined in \rcite{Trenins-JCP-2023-34108}.}
\begin{equation}
\label{eq:ZPJ-def}
Z_P^{(J)}
= \frac{1}{2J+1}\mathrm{Tr}\!\left[e^{-\beta\hat{H}}\hat{P}\mathcal{P}^{(J)}\right],
\end{equation}
where the prefactor $1/(2J+1)$ corresponds to averaging over the degenerate sublevels associated with the quantum number $M$.
Substituting \Eq{eq:proj-gamma} into \Eq{eq:delta-proj} then yields\footnote{This expression can be simplified further by exploiting the class structure of the group. If two operations $\hat{P}$ and $\hat{P}'$ belong to the same conjugacy class, then $\hat{P}' = \hat{Q}\hat{P}\hat{Q}^{-1}$ for some $\hat{Q}\in\mathcal{G}$.
Using $[\hat{H},\hat{Q}]=0$ and cyclic invariance of the trace, one finds $Z_{P'}=\mathrm{Tr}\!\left[e^{-\beta\hat{H}}\hat{Q}\hat{P}\hat{Q}^{-1}\right]=Z_P$, that is, $Z_P$ depends only on the conjugacy class of $\hat{P}$.
Since the character of $\hat{P}$ also depends only on the conjugacy class, it is sufficient to evaluate only one symmetrized partition function per class, which can be particularly advantageous for nontrivial conjugacy classes with many elements.}, again for $\beta\rightarrow\infty$,
\begin{equation}
\label{eq:delta-ZPJ}
e^{-\beta\Delta_{\Gamma\Gamma'}^{JJ'}}
= \frac{\sum_{\hat{P}\in\mathcal{G}} \chi^{(\Gamma)*}_P\,Z_P^{(J)}}
       {\sum_{\hat{P}\in\mathcal{G}} \chi^{(\Gamma')*}_P\,Z_P^{(J')}}.
\end{equation}
Hence, once the set of symmetrized partition functions $\{Z_P^{(J)}\}$ is known at sufficiently low temperature, the tunneling splitting between $(\Gamma,J)$ and $(\Gamma',J')$ can be obtained by taking the logarithm of \Eq{eq:delta-ZPJ} and dividing by $-\beta$.

Equations~\eqref{eq:ZPJ-def}  and \eqref{eq:delta-ZPJ} provide the formal foundation for computing tunneling splittings from symmetrized and rotationally projected partition functions.
It serves as the common starting point for the recently established PIMD-TI approach with an Eckart spring\cite{Trenins-JCP-2023-34108,Zupan-ExactTunnelingSplittings-2026} and for the PIHMC-EBP method introduced in the following sections.
It is worth noting that the explicit projection onto a chosen $J$ removes contamination from other rotational manifolds and, at the same time, enables the computation of tunneling splittings for both rotational ground and rotationally excited states.\cite{Zupan-ExactTunnelingSplittings-2026}
By contrast, in earlier PIMD studies targeting ground-state tunneling splittings without such a $J$-projection, rotational excitations were typically suppressed by taking $\beta$ sufficiently large, which not only significantly increases computational cost but, more problematically, often leads to unreliable splitting values.\cite{Matyus-JCP-2016-114108,Matyus-JCP-2016-114109,Vaillant-JCP-2018-234102,Vaillant-JPCL-2019-7300}

\subsection{Ring polymer with the Eckart spring}\label{sec:rp}
We now reformulate the $J$-projected symmetrized partition functions introduced above using discretized path integrals.
Inserting %
resolutions of identity and thereby introducing $N$ replicas (``beads'') of the system, \Eq{eq:ZPJ-def} can be reexpressed as\cite{Trenins-JCP-2023-34108,Zupan-ExactTunnelingSplittings-2026}
\begin{equation}
\begin{split}
Z_P^{(J)}
&= 
\frac{A_N}{8\pi^2}\int \mathrm{d}\mathbf{\Omega}\,
\chi^{(J)}(\mathbf{\Omega}) \\
&\quad\times\int \mathrm{d}\bm{q}
\exp\!\left[-\beta_N U_{P}(\bm{q},\mathbf{\Omega})\right].
\label{eq:ZPJ-PI}
\end{split}
\end{equation}
Here, $\beta_N=\beta/N$.
The prefactor $A_N$ is given by $A_N=\prod_{k=1}^{f}\left(m_k\omega_N/2\pi\hbar\right)^{N/2}$, where $f$ is the number of nuclear degrees of freedom (equal to 3 times the number of atoms, $N_\mathrm{at}$),  $m_k$ is the mass associated with the $k$-th degree of freedom, and $\omega_N=1/\beta_N\hbar$.
The ring-polymer coordinates are collected as
$\bm{q}=\{\mathbf{q}^{(1)},\dots,\mathbf{q}^{(N)}\}$,
where $\mathbf{q}^{(i)}=\{q^{(i)}_k\}_{k=1}^f$ is the Cartesian coordinate vector of bead $i$.
The ring-polymer potential entering \Eq{eq:ZPJ-PI} is decomposed as
\begin{equation}
U_{P}(\bm{q},\mathbf{\Omega})=U^{\mathrm{(open)}}(\bm{q})+U_{P}^{\mathrm{(bc)}}(\bm{q},\mathbf{\Omega}),
\label{eq:UNP}
\end{equation}
with
\begin{equation}
\begin{split}
U^{\mathrm{(open)}}(\bm{q})=&\sum_{i=1}^{N-1}\sum_{k=1}^f\frac{m_k\omega_N^2}{2}(q_{k}^{(i+1)}-q_{k}^{(i)})^2 %
+\sum_{i=1}^{N}V(\mathbf{q}^{(i)}),
\end{split}
\end{equation}
and
\begin{equation}
\label{eq:VNP}
U_{P}^{\mathrm{(bc)}}(\bm{q},\mathbf{\Omega})=\sum_{k=1}^f\frac{m_k\omega_N^2}{2}\left[\hat{P}^{-1}\mathbf{q}^{(N)}-\hat{R}(\mathbf{\Omega})\mathbf{q}^{(1)}\right]_k^2.
\end{equation}
Here, $V(\mathbf{q}^{(i)})$ is the physical potential energy evaluated at the $i$-th bead.
The first term $U^{\mathrm{(open)}}$ describes an ``open-chain" polymer in which neighboring beads $i$ and $i+1$ are connected by harmonic springs with the frequency $\omega_N$ and each bead samples the underlying physical potential.
The second term $U_{P}^{\mathrm{(bc)}}$, where bc denotes the boundary condition, closes the chain by introducing a spring between bead 1 and bead $N$, but after bead $N$ has been transformed by $\hat{P}^{-1}$ and bead 1 rotated by $\hat{R}(\mathbf{\Omega})$.
In other words, it enforces the boundary condition appropriate to the symmetrized partition function.
When $\hat{P}$ is the identity and no overall rotation is applied, $U_{P}^{\mathrm{(bc)}}$ reduces to the usual periodic closure, and $U_{P}$ recovers the standard ring-polymer potential.

A key distinction from the usual discretized path integral is the additional integration over $\mathbf{\Omega}$, which implements the exact projection onto a given total angular momentum $J$.
In a PIMD or PIHMC simulation, this could, in principle, be handled directly by treating $\mathbf{\Omega}$ as additional dynamical variables, assigning conjugate momenta, and evolving $(\bm{q},\mathbf{\Omega})$ jointly.
However, in this work, we will employ an alternative treatment, which has been proposed and used in previous works\cite{Trenins-JCP-2023-34108,Zupan-ExactTunnelingSplittings-2026,Baumann-MP-2025-2474202,Shen-JCP-2025-154307} and will be briefly described in what follows.

For fixed bead configurations $\mathbf{q}^{(1)}$ and $\mathbf{q}^{(N)}$,
the integrand $\exp(-\beta_N U_{P}^{\mathrm{(bc)}})$ is a sharply peaked function of $\mathbf{\Omega}$ around an optimal rotation
$\widetilde{\mathbf{\Omega}}$, which minimizes the mass-weighted mean-square deviation between $\hat{P}^{-1}\mathbf{q}^{(N)}$ and $\hat{R}(\widetilde{\mathbf{\Omega}})\mathbf{q}^{(1)}$.  The optimal rotation can be shown to be equivalent to the Eckart rotation that maximally decouples overall rotation from internal vibrations, and it can be computed efficiently, for example,  using a quaternion-based algorithm.\cite{Krasnoshchekov-JCP-2014-154104}

Expanding $U_{P}(\bm{q},\mathbf{\Omega})$ to second order around $\widetilde{\mathbf{\Omega}}$ and performing the angular integration by steepest descent then recasts \Eq{eq:ZPJ-PI} into\cite{Trenins-JCP-2023-34108,Zupan-ExactTunnelingSplittings-2026}
\begin{equation}
Z_P^{(J)}= A_N\int\mathrm{d}\bm{q}\,u_P^{(J)}(\bm{q})\,
e^{-\beta_N\widetilde{U}_{P}(\bm{q})},
\label{eq:ZPJ-eckart}
\end{equation}
where the new ring-polymer potential $\widetilde{U}_{P}(\bm{q})$ has the following form:
\begin{equation}\label{eq:up-tilde}
\widetilde{U}_{P}(\bm{q}) = U^{\mathrm{(open)}}(\bm{q}) + \widetilde{U}_{P}^{\mathrm{(bc)}}(\bm{q}),
\end{equation}
with the ``Eckart-spring'' term $\widetilde{U}_{P}^{\mathrm{(bc)}}(\bm{q})$ given by
\begin{equation}
\widetilde{U}_{P}^{\mathrm{(bc)}}(\bm{q})=
\sum_{k=1}^f\frac{m_k\omega_N^2}{2}\left[\hat{P}^{-1}\mathbf{q}^{(N)}-\hat{R}(\widetilde{\mathbf{\Omega}})\mathbf{q}^{(1)}\right]_k^2.
\label{eq:VNP-eckart}
\end{equation}
As can be seen, $\widetilde{U}_{P}^{\mathrm{(bc)}}(\bm{q})$ has no dependence on $\mathbf{\Omega}$ but only on $\widetilde{\mathbf{\Omega}}$, which is an implicit function of $\mathbf{q}^{(1)}$ and $\hat{P}^{-1}\mathbf{q}^{(N)}$.
The forces required by PIMD or PIHMC can be evaluated analytically.\cite{Trenins-JCP-2023-34108,Zupan-ExactTunnelingSplittings-2026}
The prefactor $u_P^{(J)}(\bm{q})$ is given by
\begin{equation}
\begin{split}
u_P^{(J)}(\bm{q})&=\frac{1}{8\pi^2}
\chi^{(J)}(\mathbf{\widetilde{\Omega}})
\left(\frac{2\pi}{\beta_N\omega_N^2}\right)^{3/2} \\
&\quad\times\det\mathbf{\Theta}\big(\hat{P}^{-1}\mathbf{q}^{(N)},\hat{R}(\widetilde{\mathbf{\Omega}})\mathbf{q}^{(1)}\big)^{-1/2}.
\label{eq:uPJ}
\end{split}
\end{equation}
Here, $\mathbf{\Theta}$ is a $3\times 3$ matrix defined for two Cartesian nuclear configurations $\mathbf{q}$ and $\mathbf{q}'$ as
\begin{equation}
\left[\mathbf{\Theta}(\mathbf{q},\mathbf{q}')\right]_{\mu\nu}=
\sum_{a=1}^{N_\mathrm{at}}m_a\left[(\mathbf{q}_a \cdot \mathbf{q}_{a}^\prime)\delta_{\mu\nu}-q_{a,\mu}q_{a,\nu}^\prime\right],
\label{eq:theta}
\end{equation}
where the index $a$ labels atoms, $\mathbf{q}_a=(q_{a,x},q_{a,y},q_{a,z})$ denotes the Cartesian coordinates of the atom $a$ with mass $m_a$, and $\mu,\nu\in\{x,y,z\}$.
In this expression, we adopt an atom-wise notation for $\mathbf{q}$ and $\mathbf{q}'$, both of which are single-bead configurations and should not be confused with the full ring-polymer configuration $\bm{q}$.
The Eckart-frame alignment between bead 1 and bead $N$ is specified by the optimal rotation $\widetilde{\mathbf{\Omega}}$, and the associated steepest-descent prefactor accounts for angular fluctuations about this alignment through the matrix $\mathbf{\Theta}$.

Unlike semiclassical instanton theory, where a steepest-descent approximation is carried out in all degrees of freedom\cite{Richardson-JCP-2011-54109,Richardson-JCP-2011-124109,Richardson-IRPC-2018-171},
the steepest-descent integration used to obtain \Eq{eq:ZPJ-eckart} introduces an error of order $\mathcal{O}(1/N)$, which vanishes in the limit of infinitely many beads.
Although this scaling is formally less favorable than the $\mathcal{O}(1/N^2)$ convergence of a symmetric Trotter factorization, in practice it does not noticeably degrade the convergence with respect to $N$ in tunneling-splitting calculations.

The effective ring-polymer potential $\widetilde{U}_{P}$ depends only on the symmetry operation $\hat{P}$ but not $J$, while the prefactor $u_P^{(J)}(\bm{q})$ depends on both.
As such, a single simulation sampling the ensemble defined by $\widetilde{U}_{P}$ provides access to symmetrized partition functions for multiple $J$ values.\cite{Zupan-ExactTunnelingSplittings-2026} %

The tunneling-splitting formulas Eqs.~\eqref{eq:delta-ZPJ} require only ratios of partition functions.
Using \Eq{eq:ZPJ-eckart}, we can express the ratio as
\begin{equation}
\frac{Z_{P}^{(J)}}{Z_{P'}^{(J')}}=\frac{\langle u_{P}^{(J)}(\bm{q})\rangle_P}{\langle u_{P'}^{(J')}(\bm{q})\rangle_{P'}}\,e^{-\beta_N\Delta F},
\end{equation}
where $\langle\cdots\rangle_P$ denotes an ensemble average with respect to $\widetilde{U}_{P}$, and $\Delta F$ is the free-energy difference between the two effective ring-polymer potentials, which is defined via
\begin{equation}
e^{-\beta_N\Delta F}=
\frac{\int\mathrm{d}\bm{q}\,e^{-\beta_N\widetilde{U}_{P}(\bm{q})}}
{\int\mathrm{d}\bm{q}\,e^{-\beta_N\widetilde{U}_{P'}(\bm{q})}}.
\label{eq:exp-dF}
\end{equation}

\subsection{Enveloping bridging potentials}\label{sec:ebp}
In practice, a single PIMD or PIHMC simulation samples configurations according to one specific ring-polymer potential.
As a consequence, direct evaluation of the free-energy difference between two such potentials is difficult whenever the corresponding ensembles have poor overlap, which is typically the case for our tunneling problems.
Previous PIMD-based approaches have therefore relied on TI to compute these free-energy differences,  inheriting the practical complications discussed earlier.
In this subsection, we introduce the concept of EBP\cite{Bennett-JCP-1976-245}, which can be used instead of TI to provide a more robust and efficient framework for obtaining the required free-energy information.

\begin{figure*}[htbp]
    \centering
    \includegraphics[width=0.98\linewidth]{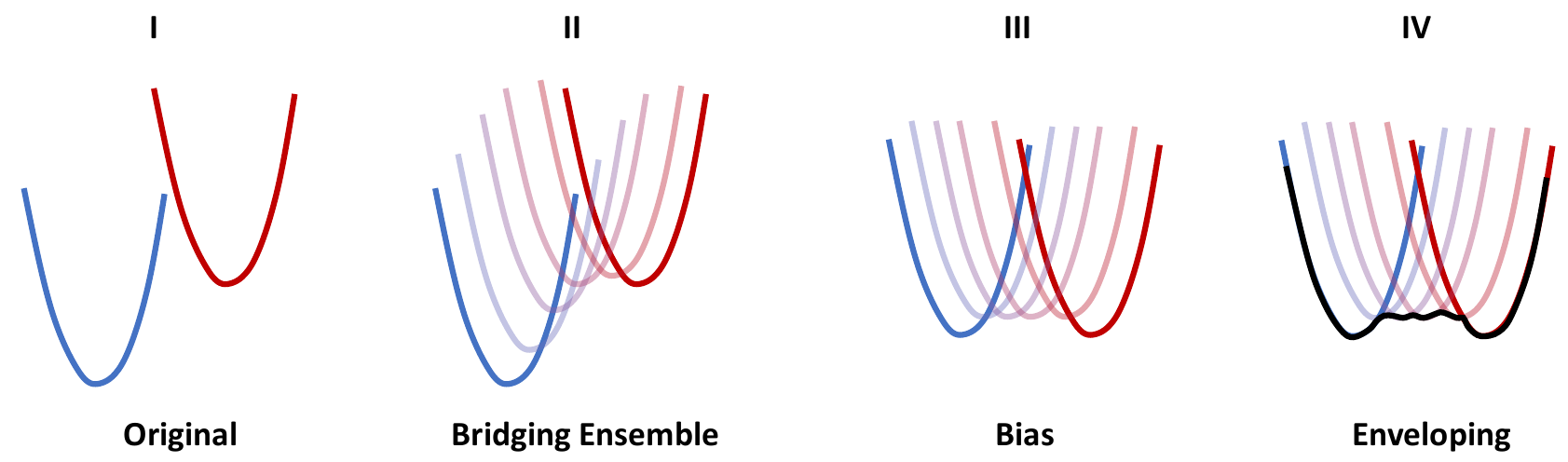}
    \caption{Schematic illustration of the stepwise construction of an EBP connecting target potentials.}
    \label{fig:ebp}
\end{figure*}

\subsubsection{EBP formulation}
The central idea is to construct a smooth thermodynamic path that connects two or more target potentials of interest, and then to augment this path with a set of appropriately biased bridging potentials.
These potentials are combined into a single effective potential that ``envelops" the targets while removing barriers along the chosen path in configuration space, as illustrated schematically in \Fig{fig:ebp}.
Consequently, a single PIMD or PIHMC simulation performed with this enveloping potential can efficiently sample configurations associated with all relevant regions, enabling direct estimation of the free-energy profile without numerical integration over separate intermediate states.

To make this more concrete, consider the two target potentials appearing on the right-hand-side of \Eq{eq:exp-dF}.
We introduce a parameter $\lambda\in[0,1]$ which describes a linearly interpolated thermodynamic path connecting the two targets via intermediate ring-polymer potentials
\begin{equation}
U_\mathrm{m}(\bm{q};\lambda)=(1-\lambda)\widetilde{U}_{P}(\bm{q})+\lambda\widetilde{U}_{P'}(\bm{q}).
\end{equation}
In the spirit of bridging potentials, we select a discrete set of parameters $\{\lambda_l\}$ along this path and associate a bias $\varepsilon_l$ with each one.
The EBP is then expressed as
\begin{equation}
U_\mathrm{EBP}(\bm{q})=-\frac{1}{\beta_N}\ln\left[\sum_l e^{-\beta_N\left(U_\mathrm{m}(\bm{q};\lambda_l)-\varepsilon_l\right)}\right].
\label{eq:VEBP}
\end{equation}
Alternative choices of thermodynamic paths other than the linear interpolation could also be employed.
In the present work, however, we focus on this simplest form, which we find to be sufficient for efficient tunneling-splitting calculations.

Although \Eq{eq:VEBP} involves a sum over many intermediate states, evaluating $U_\mathrm{EBP}(\bm{q})$ is as easy as evaluating the two end-state potentials $\widetilde{U}_{P}$ and $\widetilde{U}_{P'}$.
This is because these two potentials differ only in the Eckart-spring boundary term, while all remaining contributions are identical.
We therefore compute the common open-chain contribution only once per configuration, together with the two endpoint Eckart-spring terms, and obtain any intermediate state along the linear path by interpolating only the boundary term.
In particular, the boundary contribution for an intermediate state is
\begin{equation}
U_\mathrm{m}^{\mathrm{(bc)}}(\bm{q};\lambda)=(1-\lambda)\widetilde{U}_{P}^{\mathrm{(bc)}}(\bm{q})+\lambda\widetilde{U}_{P'}^{\mathrm{(bc)}}(\bm{q}).
\end{equation}
Since all intermediate states share the same open-chain part, this contribution can be factored out of \Eq{eq:VEBP}, which gives
\begin{equation}
\label{eq:VEBP-bc}
\begin{split}
U_\mathrm{EBP}(\bm{q})=&-\frac{1}{\beta_N}\ln\left[\sum_l e^{-\beta_N\left(U_\mathrm{m}^{\mathrm{(bc)}}(\bm{q};\lambda_l)-\varepsilon_l\right)}\right] \\
&+U^{\mathrm{(open)}}(\bm{q}).
\end{split}
\end{equation}
This form makes it clear that the evaluation of $U_\mathrm{EBP}$ requires no additional calculations of the physical potential beyond those already needed for the corresponding ring polymer, since the dependence on $\lambda$ enters only through the boundary term.
The associated  force required by PIMD or PIHMC is obtained straightforwardly by applying the chain rule to \Eq{eq:VEBP-bc}.

Intuitively, the quantity inside the logarithm of \Eq{eq:VEBP} is a sum of thermodynamic weights associated with the biased bridging potentials $U_\mathrm{m}(\bm{q};\lambda_l)-\varepsilon_l$.
For any given configuration $\bm{q}$, this sum is necessarily larger than the contribution from any individual term.
Interpreting the entire sum as the Boltzmann factor of an effective potential implies that $U_\mathrm{EBP}(\bm{q})$  is always less than each of the biased bridging potentials at that configuration.
Consequently, $U_\mathrm{EBP}(\bm{q})$ forms a continuous envelope that lies below and smoothly connects low-energy regions of biased potentials distributed along the thermodynamic path.
With a suitable choice of $\{\lambda_l\}$ and $\{\varepsilon_l\}$, the envelope can be made to cover the target potentials while rendering the path between them approximately barrierless.
Under this enveloping potential, a single simulation can then explore all important regions, providing direct evaluation of the free-energy profile along the whole path.
Moreover, observables (for example, $O$) under any constituent bridging potential $U_\mathrm{m}(\bm{q};\lambda)$ (including the end states) can be obtained from the same trajectory by reweighting according to
\begin{equation}
\langle O(\bm{q})\rangle_{\lambda} = 
\frac{\left\langle O(\bm{q})e^{-\beta_N\left[U_\mathrm{m}(\bm{q};\lambda)-U_{\mathrm{EBP}}(\bm{q})\right]}\right\rangle_{\mathrm{EBP}}}
{\left\langle e^{-\beta_N\left[U_\mathrm{m}(\bm{q};\lambda)-U_{\mathrm{EBP}}(\bm{q})\right]}\right\rangle_{\mathrm{EBP}}},
\end{equation}
where $\left\langle\cdot\right\rangle_\mathrm{EBP}$ denotes an ensemble average with respect to $U_{\mathrm{EBP}}$.
Note that although the above equation is rigorous for any EBP, the numerical convergence strongly depends on the construction of the grid and the choice of bias values.

The underlying idea of using an enveloping potential constructed from multiple biased intermediate states is conceptually straightforward and has appeared in various forms since the 1970s.\cite{Bennett-JCP-1976-245,Han-PLA-1992-28}
Related strategies have been revisited and generalized in recent years\cite{Lu-PRE-2004-57702,Christ-JCP-2007-184110,Christ-JCP-2008-174112,Christ-JCTC-2009-276,Mori-TJoPCB-2014-8210,Sidler-JCP-2016-154114,Koenig-JCIM-2020-5407,Koenig-JCTC-2021-5805,Lin-JCTC-2026-818}, notably in biomolecular simulations.
Nevertheless, in practice, selecting suitable bridging grids and determining optimal biases is highly nontrivial.
In earlier studies, the grid is often chosen in a simple uniform manner, while the corresponding bias parameters are refined iteratively through repeated long simulations.
This strategy is inefficient and rarely optimal, because the free-energy profile along a thermodynamic path is generally neither flat nor simple, but may contain steep barriers and complex structures.
As emphasized by Bennett\cite{Bennett-JCP-1976-245}, reliably selecting a good distribution of $\lambda$ grid points and biases is almost equivalent to knowing the full free-energy landscape in advance.

Fortunately, in ring-polymer-based calculations of tunneling splittings, we can leverage RPI as a cost-effective guide for this design.
In practice, RPI calculations are orders of magnitude cheaper than a fully converged PIMD simulation, while remaining quantitatively reliable often within about 20\% of the correct result.\cite{Richardson-PCCP-2017-966,Sahu-JCC-2021-210,Kaeser-JCTC-2022-6840,Nandi-JACS-2023-9655}
We therefore propose to pre-scan the chosen thermodynamic path by fast instanton calculations at a set of provisional $\lambda$ values
and use this information to determine an intelligent distribution of grid points and to set reasonable biases.
This guidance makes the construction of an approximately barrierless EBP practical and reliable, thereby enabling a numerically efficient path-integral simulation.

\subsubsection{Construction guided by RPI}
Let us consider the ring-polymer potential $U_\mathrm{m}(\bm{q};\lambda)$ along a prescribed thermodynamic path parameterized by $\lambda$.
In RPI, one first locates the global minimum $\tilde{\bm{q}}$ (which we assume to be unique for simplicity) of $U_\mathrm{m}(\bm{q};\lambda)$ by optimizing the ring-polymer configuration.
The corresponding partition function $Z_\mathrm{m}(\lambda)$ is then estimated by a steepest-descent integration around $\tilde{\bm{q}}$.
When the kink-induced zero mode is absent, this approximation gives\cite{Richardson-JCP-2011-54109,Richardson-JCP-2011-124109,Richardson-IRPC-2018-171}
\begin{equation}
\begin{split}
Z_\mathrm{m}(\lambda)&=A_N\int\mathrm{d}\bm{q}\,
e^{-\beta_NU_\mathrm{m}(\bm{q};\lambda)} \\
&\simeq A_N Z_\mathrm{free}\left(\prod_{\kappa}\frac{2\pi}{\beta_N\gamma_{\kappa}}\right)^{1/2}e^{-\beta_NU_\mathrm{m}(\tilde{\bm{q}};\lambda)},
\end{split}
\end{equation}
where $\{\gamma_{\kappa}\}$ are the %
eigenvalues of the %
Hessian of $U_\mathrm{m}(\bm{q};\lambda)$ at $\tilde{\bm{q}}$, excluding the overall translational and rotational normal modes of the ring polymer, which are accounted for in $Z_\mathrm{free}$.
If a zero-frequency mode due to a kink exists, that mode must be treated analytically, leading to
\begin{equation}
Z_\mathrm{m}(\lambda)\simeq
\beta\hbar|\mathbf{v}_0| A_N Z_\mathrm{free}\left({\prod_{\kappa}}^\prime\frac{2\pi}{\beta_N\gamma_{\kappa}}\right)^{1/2}e^{-\beta_NU_\mathrm{m}(\tilde{\bm{q}};\lambda)},
\end{equation}
where the prime on the product symbol indicates omission of the zero mode and
\begin{equation}
|\mathbf{v}_0|=\sqrt{\sum_{i=2}^{N-1}\sum_{k=1}^f\left(\frac{\tilde{q}_k^{(i+1)}-\tilde{q}_k^{(i-1)}}{2\beta_N\hbar}\right)^2}
\end{equation}
is the normalization factor associated with the zero mode.
These expressions provide fast and reasonably accurate estimates of both the Euclidean action $\beta_NU_\mathrm{m}(\tilde{\bm{q}};\lambda)$ (in units of $\hbar$) and the corresponding $Z_\mathrm{m}(\lambda)$ at the selected $\lambda$ values.

In practice, we choose a set of preliminary $\lambda$ grid points\footnote{The chosen $\lambda$ grid points are 0, 0.0001, 0.001, 0.005, 0.01, 0.02, 0.05, 0.1, 0.15, 0.2, 0.25, 0.3, 0.35, 0.4, 0.5, 0.6, 0.7, 0.8, 0.9, 0.95, 0.98, 0.99, 0.995, 1. Here $\lambda=0$ and 1 correspond to the lower- and higher-action end states, respectively.} (denser near $\lambda=0$ and $\lambda=1$, where free-energy variations are typically more pronounced) and evaluate $\beta_NU_\mathrm{m}(\tilde{\bm{q}};\lambda)$ by RPI at those points.
Then, we fit the profile of $\beta_NU_\mathrm{m}(\tilde{\bm{q}};\lambda)$ with a flexible double-hill model,
\begin{equation}
y(\lambda)=y_0+a_1\frac{\lambda}{K_1^{c_1}+\lambda^{c_1}}-a_2\frac{\lambda}{K_2^{c_2}+\lambda^{c_2}},
\end{equation}
where $y_0$, $a_1$, $a_2$, $c_1$, $c_2$, $K_1$, and $K_2$ are fitting parameters.
We have found that this functional form reproduces the characteristic shape observed for most Euclidean action profiles and free-energy profiles along thermodynamic paths connecting symmetrized partition functions.
Figures~S4-S7 present the fitted curves together with the original RPI results for the molecular systems studied in \Sec{sec3}.
From the fitted curve, we construct an adaptive grid by assigning each $\lambda$ a weight proportional to the gradient of the fitted curve, normalizing to obtain a density and its cumulative distribution, and then choosing grid points from the inverse cumulative distribution function.
The resulting grid allocates more points to rapidly varying regions and fewer to flat regions.
From our experience, targeting a grid spacing of $\Delta y$ in the range of 0.1 to 0.2 yields sufficient overlap without excessive redundancy.

The same double-hill model is applied to fit the dimensionless bias $\tilde{\epsilon}_l$ calculated according to
\begin{equation}
\tilde{\epsilon}_l\equiv\beta_N\varepsilon_l=-\ln\left[\tanh\left(\frac{Z_\mathrm{m}(\lambda_l)}{Z_\mathrm{ref}}\right)\right],
\end{equation}
 where $Z_\mathrm{ref}$ denotes the symmetrized partition function of the end state with the smaller RPI action, and both $Z_\mathrm{m}(\lambda_l)$ and $Z_\mathrm{ref}$ are evaluated by RPI.
This expression follows the resummation strategy used in instanton-based tunneling-splitting theory justified for the double-well tunneling case\cite{Benderskiui-ChemicalDynamicsLow-1994,Richardson-JCP-2011-54109} and generally provides a more reliable free-energy estimate than a raw ratio of partition functions, including in multi-well cases.

Moreover, this construction can be further refined by a short preliminary PIMD or PIHMC simulation using the RPI-based estimates as initial guesses for the biases. The resulting trajectories provide a more accurate estimate of the underlying free-energy profile along the thermodynamic path, which can then be used directly to update the bias parameters.
Since the biases from RPI are already close to their optimal values, only relatively short simulations are required to obtain a near-optimal set of bias values.

\subsubsection{Composite end state}
The construction defined in \Eq{eq:VEBP} is not restricted to a two-state setting but also applies directly to multi-well tunneling problems where multiple end states are involved.
In this case, one may define a set of thermodynamic paths, each connecting a chosen pair of end states, and form a connected network that links all end states without gaps.
The selection of this network is flexible.
A practical choice is to prefer paths with smaller free-energy variations while maintaining overall connectivity.
This information can be easily obtained from the preliminary RPI scans described above.

In some situations, however, it is unnecessary to introduce bridging states between certain end states.
One common example occurs when the corresponding symmetry operations belong to the same conjugacy class.
In that case, the corresponding ring-polymer ensembles are related by symmetry, so that they play an equivalent role in the tunneling-splitting expressions.
Another situation arises when two end states have sufficiently high configurational overlap so that direct sampling on either end state is already effective.
For these cases, we group the relevant end states $\{\widetilde{U}_{P}\}$ into a composite set $\mathcal{S}$ and treat it as a single effective end state associated with the following potential:
\begin{equation}
U_\mathcal{S}(\bm{q})=-\frac{1}{\beta_N}\ln
\left[\sum_{P\in\mathcal{S}}e^{-\beta_N\widetilde{U}_{P}(\bm{q})}\right].
\end{equation}
This reduction is particularly advantageous for large symmetry groups with conjugacy classes of high order because it decreases the number of distinct end states and therefore reduces the number of thermodynamic paths that need to be considered.
This scheme will be illustrated for the water dimer in \Sec{sec4-water-dimer}.

\subsubsection{Windowing}\label{sec:windowing}

The EBP scheme described above can, in principle, include multiple target ring-polymer potentials within a single barrierless effective landscape.
This implies that a single path-integral simulation performed under this global enveloping potential would then provide direct estimates of all required free-energy differences between the target distributions.
For large molecular symmetry groups this appears particularly attractive, since it suggests that the whole tunneling-splitting pattern could be computed from one simulation, whereas with TI, one would require a separately converged TI for each conjugacy class.

In practice, however, this fully global construction is rarely optimal.
Sampling along a barrier-free thermodynamic path proceeds through an essentially diffusive motion in the configuration space, so that the typical distance explored along the path grows only with the square root of simulation time.
When the path is short, configurations travel efficiently between endpoints, and a single enveloping potential is adequate.
However, when the path becomes long, as in tunneling problems with high barriers or when several distinct thermodynamic paths are concatenated, the diffusive motion between distant regions becomes slow, and the autocorrelation time, \textit{i.e.}, the time required to generate effectively independent configurations, increases significantly.
The resulting loss of statistical efficiency outweighs the conceptual simplicity of a single simulation.

\begin{figure}[htbp]
    \centering
    \includegraphics[width=0.98\linewidth]{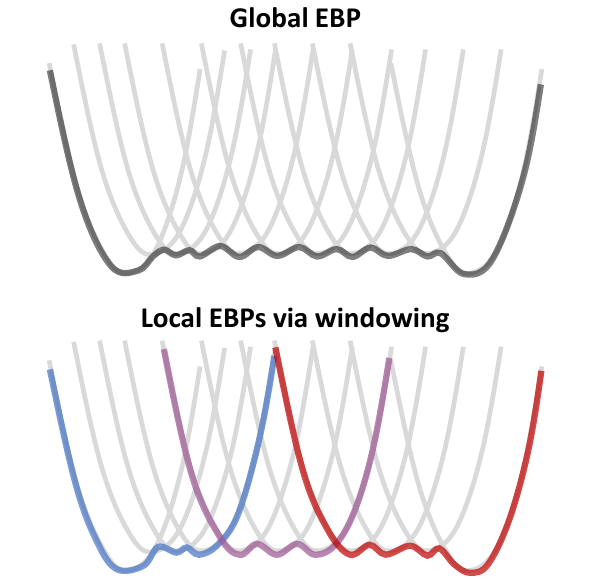}
    \caption{Schematic illustration of windowing a global EBP into overlapping local EBPs.}
    \label{fig:window}
\end{figure}

A more effective strategy is therefore to partition the thermodynamic path into a set of overlapping windows once the grid points and corresponding biases have been chosen, as illustrated in \Fig{fig:window}.
Each window spans a contiguous segment of the path and is chosen to be of moderate length, long enough to avoid an excessive number of windows but not so long that diffusion becomes slow and autocorrelation times increase.
In practice, the resulting number of windows is typically far smaller than the number of quadrature points required in TI.
In contrast to TI, windowing introduces no systematic error but only affects sampling efficiency.
Independent path-integral simulations are then performed in these windows using the sampling scheme introduced in \Sec{sec:pihmc}.
The collection of configurations from all windows can be combined in a statistically optimal manner, as described in \Sec{sec:mbar}.

To promote sufficient sampling of the end states, we apply an additional bias to the end-state terms in the local EBP.
Let $N_{p}^{(w)}$ denote the number of constituent potentials included in window $w$.
If the window contains a single end state, we add $\frac{1}{\beta_N}\ln\!\big(N_{p}^{(w)}-1\big)$ to the bias parameter of that end-state term, so that it has approximately one half of the total weight in that window when the remaining terms have comparable effective weights.
If a window contains multiple end states, we apply an analogous shift to each end state so that their combined weight is approximately one half.

\subsection{Path-integral hybrid Monte Carlo}\label{sec:pihmc}
\subsubsection{General description}
Starting from the window ensembles introduced above, we now consider a generic window partition function written as a configurational integral over the corresponding ring-polymer potential $U^{(w)}$.
We then introduce fictitious momenta
$\bm{p}=\{\mathbf{p}^{(1)},\dots,\mathbf{p}^{(N)}\}$ with
$\mathbf{p}^{(i)}=\{p^{(i)}_k\}_{k=1}^f$ for each bead.
This yields the phase-space representation
\begin{equation}
Z^{(w)}=(2\pi\hbar)^{-Nf}\int\mathrm{d}\bm{p}\int\mathrm{d}\bm{q}\,
e^{-\beta_N H^{(w)}(\bm{p},\bm{q})},
\label{eq:ZPJ-pihmc}
\end{equation}
where the ring-polymer Hamiltonian,
\begin{equation}
H^{(w)}(\bm{p},\bm{q})
=\sum_{i=1}^{N}\sum_{k=1}^f\frac{(p_k^{(i)})^2}{2m_k}+U^{(w)}(\bm{q}),
\label{eq:HNP}
\end{equation}
consists of the fictitious kinetic energy and the window potential $U^{(w)}(\bm{q})$ defined by the local EBP in window $w$.

This Hamiltonian is isomorphic to a classical system in an extended phase space, in which $N$ replicas are coupled by harmonic springs.
Ensemble averages with respect to this distribution can therefore be generated either by thermostatted molecular dynamics, leading to conventional PIMD, or by HMC\@.
As explained earlier, since the use of a finite time step in PIMD introduces integration errors that may bias the sampling unless carefully converged, we adopt HMC in the present work to obtain unbiased configurational sampling while retaining good efficiency.

HMC, also called Hamiltonian Monte Carlo\cite{Duane-PLB-1987-216}, constructs Markov-chain updates by combining short molecular-dynamics trajectories with a Metropolis acceptance step.
The extension of HMC to imaginary-time path integrals is often referred to as PIHMC\@.\cite{Tuckerman-JCP-1993-2796}
At the beginning of each Monte Carlo move, all bead momenta are independently resampled from the normal distribution with standard deviation $\sqrt{m_k/\beta_N}$  according to the kinetic term in $H^{(w)}(\bm{p},\bm{q})$.
The coordinates and momenta are then propagated for a fixed number of steps $L$, thus generating a trial phase-space point $(\bm{p}',\bm{q}')$.
This proposal is accepted with probability
\begin{equation}
P_\mathrm{acc}=\mathrm{min}\left[1,\exp\big(-\beta_N[H^{(w)}(\bm{p}',\bm{q}')-H^{(w)}(\bm{p},\bm{q})]\big)\right],
\end{equation}
otherwise the old configuration is retained.
Exact sampling is guaranteed by numerically time-reversible and volume-preserving integrators.
The Metropolis step eliminates time step bias as long as an adequate acceptance rate is maintained.

In principle, a wide variety of integrators with the required properties can be used in HMC.
A common and robust choice is the velocity-Verlet scheme, which we adopt in this work.
Compared with the classical system, however, the ring-polymer potential contains harmonic spring terms that couple neighboring beads.
For a large number of beads, these springs are stiff, which severely restricts the size of the time step that can be used while keeping the Metropolis rejection rate under control.
To overcome this issue, we employ a multiple-time-step (MTS) integration strategy analogous to that used in previous PIMD-TI works with the Eckart spring formulation.\footnote{Note that integrators based on ring-polymer normal modes or Cayley transforms are not directly applicable to the case with an Eckart spring.}
The total force is decomposed into a fast component arising from the stiff spring interactions and a slow component arising from the physical potential.
The fast forces are integrated with a small inner time step $\delta t$, while the slow forces are updated less frequently with a larger outer time step $\Delta t$, within a time-reversible splitting scheme.
This separation avoids frequent evaluations of the physical potential and its forces, which are typically the computational bottleneck of the simulation.

\subsubsection{Nonlocal permutation update }
When PIHMC is applied to window potentials $U^{(w)}$, the resulting scheme is, in principle, sufficient to evaluate the required free-energy differences.
In practice, however, a severe sampling problem arises in a window when an instanton-like kink structure is present only in the vicinity of an end state associated with a nontrivial symmetry operation, and is absent elsewhere within that window.
In this situation, the PIHMC trajectories exhibit extremely long autocorrelation times, so that even very long simulations yield only a small number of statistically independent samples.
As a result, the statistical uncertainty in the free-energy estimates becomes prohibitively large, especially at low temperatures and for large numbers of beads.

\begin{figure*}[htbp]
    \centering
    \includegraphics[width=0.98\linewidth]{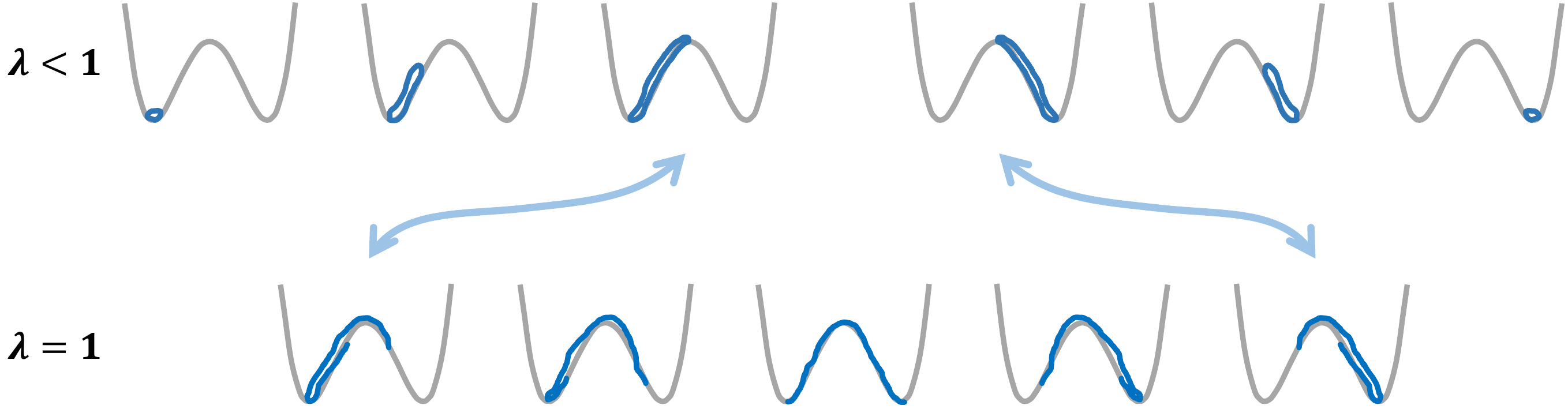}
    \caption{Schematic illustration of kink-induced trapping using a symmetric double-well model. The two relevant symmetry operations are the identity and the reflection between the wells, which define the $\lambda=0$ and $\lambda=1$ end states, respectively. In the upper row, for $\lambda=0$, the lowest-action configuration is a closed chain localized in either well. For $0<\lambda<1$, the reflection condition rapidly drives the ring polymer into a kinked closed-chain configuration near the barrier, where both boundary conditions can be satisfied. In the lower row, at $\lambda=1$, the closure constraint is removed and only the reflection condition remains, so a large manifold of quasi-degenerate broken-chain configurations becomes accessible. Once the system enters this manifold, return to the $\lambda<1$ closed-chain region can occur only when the two end beads of the chain happen to reconnect by diffusion.}
    \label{fig:kink-trap}
\end{figure*}

Such behavior can be understood by recalling the symmetry properties of the ring-polymer representation.
For a conventional ring polymer without symmetry operations, where the end beads are connected by a standard harmonic spring, the potential is invariant under cyclic permutations of the beads along imaginary time.
Permuting the bead indices shifts configurations along the ring without changing the bead coordinates.
The situation changes qualitatively once a nontrivial symmetry operation is imposed at the endpoints.
In that case, the first and last beads are constrained by the related symmetry operation, and the intervening beads form an instanton-like kink that connects them.
The location of the kink around the ring is associated with an approximate zero mode.\cite{Richardson-JCP-2011-54109,Lawrence-JCP-2023-14111}
Cyclic permutations that move beads from one end of the ring to the other then induce large apparent changes in the bead coordinates, while leaving the total ring-polymer potential essentially unchanged.
In other words, many distinct bead configurations correspond to the same kinked path translated along imaginary time.
Figure~\ref{fig:kink-trap} presents a schematic diagram that illustrates the kink-induced trapping mechanism using a symmetric double-well model.

This degeneracy implies that the phase-space volume associated with kinked configurations can be much larger than that associated with configurations in the barrierless central region, where no such zero mode is present.
The imbalance becomes more pronounced as the imaginary-time length and the number of beads increase.
During a PIHMC simulation, once the trajectory enters a kinked region of configuration space, it diffuses among the numerous quasi-degenerate kink realizations and returns only very rarely to the central, non-kinked region.
The integrated autocorrelation time of observables that distinguish these regions is therefore essentially the mean first-passage time for such rare returns, which can be extremely large.
This trapping mechanism explains the poor sampling efficiency in naive implementations that couple PIHMC directly with EBP in the presence of instanton-like structures.

To overcome this sampling difficulty, we introduce a special Monte Carlo update that enables the ring polymer to escape from the large kink manifold.
The idea is to construct a nonlocal move that efficiently relocates the kink along the imaginary-time contour while preserving detailed balance with respect to the target distribution.

In each such move, we first randomly choose a bond between two adjacent beads as a cut point.
This cut divides the ring polymer into a ``head" segment (where bead 1 is) and a ``tail" segment (where bead $N$ is) while preserving the original bead ordering within each segment.
We then attempt to reattach the head to the tail after a combined symmetry and rotation operation.
Specifically, for the beads in the head segment we apply, in sequence, the optimal Eckart rotation $\hat{R}(\widetilde{\mathbf{\Omega}})$, a small additional random rotation $\hat{R}(\Delta\mathbf{\Omega})$, and the symmetry operation $\hat{P}$ that induces the kink structures.
The transformed head is then concatenated to the tail, becoming the new junction of the ring.
This construction preserves the connectivity and spring structure while effectively shifting the kink configuration along the ring polymer.

To ensure microscopic reversibility and allow for a Metropolis acceptance test, the additional rotation $\hat{R}(\Delta\mathbf{\Omega})$ is parameterized by Euler angles $\Delta\mathbf{\Omega}=(\Delta\alpha,\Delta\beta,\Delta\gamma)$ drawn from an explicit probability density.
For each Euler-angle component $s\in(\Delta\alpha,\Delta\beta,\Delta\gamma)$ we use a symmetric exponential form,
\begin{equation}
\label{eq:prot}
P^\mathrm{rot}(s)=\mathcal{C}_se^{-a_0N|s|},
\end{equation}
where $a_0$ is the width of the distribution, $\mathcal{C}_s$ is the normalization factor, and  $s$ is restricted to $[-\pi,\pi]$, $[-\pi/2,\pi/2]$, and $[-\pi,\pi]$ for $s$ being $\alpha$, $\beta$, and $\gamma$, respectively.
The joint sampling probability $P_\mathrm{joint}^{\mathrm{rot}}(\Delta\mathbf{\Omega})$ is given by the product of the marginal densities in $\Delta\alpha$, $\Delta\beta$, and $\Delta\gamma$.
In the reverse move, we cut at the corresponding junction in the proposed configuration and transform the new tail beads by applying $\hat{P}^{-1}$, the inverse of a compensating rotation $\hat{R}(\Delta\mathbf{\Omega}')^{-1}$, and the inverse optimal Eckart rotation $\hat{R}(\widetilde{\mathbf{\Omega}}')$, where $\widetilde{\mathbf{\Omega}}'$ is the optimal rotation for the proposed configuration and $\Delta\mathbf{\Omega}'$ is chosen such that
\begin{equation}
\hat{R}(\Delta\mathbf{\Omega}')\hat{R}(\widetilde{\mathbf{\Omega}}')=\hat{R}(\Delta\mathbf{\Omega})\hat{R}(\widetilde{\mathbf{\Omega}})
\end{equation}
is satisfied.
With these definitions the proposal is exactly reversible, and the acceptance probability for the move follows from the Metropolis criterion,
\begin{equation}
\label{eq:pacc}
P_\mathrm{acc}=\mathrm{min}\left[1,
\frac{P_\mathrm{joint}^{\mathrm{rot}}(\Delta\mathbf{\Omega}') \exp(-\beta_N U^{(w)}(\bm{q}'))}
{P_\mathrm{joint}^{\mathrm{rot}}(\Delta\mathbf{\Omega}) \exp(-\beta_N U^{(w)}(\bm{q}))}\right],
\end{equation}
where $\bm{q}$ and $\bm{q}'$ are the ring-polymer configurations before and after the update, respectively.

The efficiency of this collective move depends on the width of the rotation distribution.
Guided by analysis of the Eckart spring contribution, we choose the dimensionless value
\begin{equation}
\label{eq:angle-width}
a_0=\frac{1}{N}\left(\beta_N\omega_N^2\right)^{1/2}\left(\det \mathbf{I}_\mathrm{m}\right)^{1/6},
\end{equation}
where $\mathbf{I}_\mathrm{m}$ is the inertia tensor of the molecular system at the minimum-energy geometry.
This choice yields efficient proposal moves with acceptance probabilities typically around one half.

In general, a given window $w$ may involve multiple symmetry operations among its constituent terms, and different operations may dominate at different times along the trajectory.
A simple strategy is therefore to randomly choose one $\hat{P}$ for each update and apply the nonlocal update described above.
However, when the number of relevant $\hat{P}$ becomes large, uniform selection is inefficient.
In this case, it is advantageous to choose $\hat{P}$ by importance sampling.
Specifically, for the current configuration $\bm{q}$ we compute, for each symmetry operation $\hat{P}$ included among the constituents of window $w$, the following weight factors: 
\begin{equation}
w_P^{(w)}(\bm{q})=e^{-\beta_N\widetilde{U}_{P}(\bm{q})},
\end{equation}
and select $\hat{P}$ stochastically according to the probability 
\begin{equation}
\pi_P(\bm{q})=\frac{w_P(\bm{q})}{\sum_Pw_P(\bm{q})}.
\end{equation}
The Metropolis acceptance probability \Eq{eq:pacc} is then replaced by
\begin{equation}
P_\mathrm{acc}=\mathrm{min}\left[1,
\frac{\pi_P(\bm{q}')P_\mathrm{joint}^{\mathrm{rot}}(\Delta\mathbf{\Omega}') \exp(-\beta_N U^{(w)}(\bm{q}'))}
{\pi_P(\bm{q})P_\mathrm{joint}^{\mathrm{rot}}(\Delta\mathbf{\Omega}) \exp(-\beta_N U^{(w)}(\bm{q}))}\right],
\end{equation}
so that each update preferentially selects symmetry operations with larger weights.

In practice, this update is highly effective.
Within the PIHMC scheme, performing roughly five of these permutation moves after each conventional MD trajectory update is sufficient to eliminate the kink-induced autocorrelation across a wide range of temperatures and bead numbers.
It should be noted that this update preserves the physical potential of each bead, since symmetry and overall rotation operations leave the molecular potential energy invariant.
As a result, only the change in the spring potential (including the Eckart spring) needs to be evaluated, and the computational overhead of these moves is negligible compared with the PIHMC trajectory propagation.

\subsubsection{Inter-bead rotation update}
The ratios between symmetrized partition functions require evaluation of the prefactors $u_P^{(J)}$, which contain the character of the Wigner $D$-matrix.
For $J=0$, the character equals unity, whereas for $J>0$ it depends on the relative orientation between the two end beads.
As a collective motion involving all beads, this end-to-end orientation typically evolves very slowly in the HMC trajectory-propagation update, which leads to long autocorrelation times in the $J>0$ prefactor estimators and degrades the sampling efficiency for tunneling splittings of rotationally excited states.
To alleviate this problem, we introduce a second specially designed nonlocal update, termed the inter-bead rotation update.

In each inter-bead rotation move, starting from a configuration $\bm{q}$, we first select a bead index $i$ at random. We then draw a random rotation angle from the probability density in \Eq{eq:prot}, with the width parameter $a_0$ set to twice the value in \Eq{eq:angle-width}.
The selected bead and all subsequent beads up to the tail are then rotated, each about its own center of mass, generating a proposed configuration $\bm{q}'$.
By construction, this update leaves all contributions to $U^{(w)}(\bm{q})$ unchanged except for a single open-chain spring term between the selected bead and its predecessor.\footnote{The boundary term in an EBP [\Eq{eq:VEBP-bc}] is invariant under a rigid rotation of beads $i,\ldots,N$ because it is defined through a logarithmic sum over biased Eckart-spring boundary contributions, and each constituent Eckart-spring term [\Eq{eq:VNP-eckart}] is itself invariant to such a rigid rotation when evaluated after optimal Eckart alignment of beads 1 and $N$.}
The corresponding Metropolis acceptance probability is therefore calculated according to
\begin{equation}
P_\mathrm{acc}=\mathrm{min}\left[1,e^{-\beta_N\Delta U(\bm{q},\bm{q}')}\right]
\end{equation}
with
\begin{equation}
\begin{split}
\Delta U(\bm{q},\bm{q}')=&\sum_{k=1}^f\frac{m_k\omega_N^2}{2}
({q_{k}'}^{(i)}-{q_{k}'}^{(i-1)})^2\\
&-\sum_{k=1}^f\frac{m_k\omega_N^2}{2}
(q_{k}^{(i)}-q_{k}^{(i-1)})^2 .
\end{split}
\end{equation}
Because evaluating this acceptance probability involves only a single spring term, the additional cost per move is negligible. One can therefore perform many inter-bead rotation moves between successive HMC trajectory-propagation updates. The number of such moves can be chosen based on the number of beads and the imaginary-time length, such that long-time correlations associated with internal relative rotational motion along the ring polymer are efficiently removed while retaining a small overall overhead.

Finally, it is worth noting that both this inter-bead rotation update and the nonlocal permutation update described above would, in principle, require a contravariant transformation of the fictitious momenta in addition to updating the coordinates. In the present PIHMC scheme, however, all bead momenta are fully resampled at the beginning of each trajectory-propagation move, which allows us to omit momentum updates in these two nonlocal moves.

\subsubsection{Reweighting across PESs}\label{sec:reweight-pes}
In some applications, one is interested in tunneling splittings and related observables on several PESs for the same molecular system.
A straightforward approach is to repeat the full PIHMC procedure separately on each PES.
While this is always possible, it is often not the most efficient option.
A key advantage of stochastic sampling is that, once configurations are generated from a reference ensemble, reweighting can directly yield results for other similar ensembles without additional sampling.
This enables a single PIHMC simulation propagated on a chosen reference PES to deliver results for multiple PESs.

Concretely speaking, we propagate PIHMC trajectories under a ring-polymer potential associated with a reference PES, denoted $V_\mathrm{ref}$.
For any target PES $V_\mathrm{tgt}$, the only PES-dependent contribution to the ring-polymer potential is the physical potential term, so that the energy difference required for reweighting is
\begin{equation}
\Delta U_\mathrm{tgt,ref}(\bm{q}) = \sum_{i=1}^N
\left[V_\mathrm{tgt}(\mathbf{q}^{(i)})-V_\mathrm{ref}(\mathbf{q}^{(i)})\right].
\end{equation}
During sampling, we therefore store, for each sampled configuration $\bm{q}$, the physical-potential contribution summed over all beads on the reference and target PESs, which is sufficient to evaluate $\Delta U_{\mathrm{tgt,ref}}(\bm q)$.
Given any observable $O(\bm{q})$ that depends only on the ring-polymer coordinates, its expectation value in the target ensemble is obtained from the reference samples by standard reweighting,
\begin{equation}
\langle O(\bm{q})\rangle_\mathrm{tgt} =
\frac{\langle O(\bm{q})e^{-\beta_N\Delta U_\mathrm{tgt,ref}(\bm{q})}\rangle_\mathrm{ref}}
{\langle e^{-\beta_N\Delta U_\mathrm{tgt,ref}(\bm{q})}\rangle_\mathrm{ref}}.
\end{equation}
Equivalently, for any stored sampled configuration $\bm{q}$, the target-PES ring-polymer potential needed for subsequent analysis can be obtained by adding $\Delta U_\mathrm{tgt,ref}(\bm{q})$ to the reference ring-polymer potential.
This provides, for each configuration, the reduced potentials on the target PES that are needed for the post-processing described in \Sec{sec:mbar}.
These data can then be used to obtain statistically optimal estimates of the free-energy differences and the prefactor averages on each PES.
In this way, the tunneling splittings for each PES can be obtained simultaneously.

This strategy offers two practical advantages for the calculation of tunneling splittings.
First, when several high-quality PESs differ only slightly, their path-integral ensembles typically exhibit substantial overlap, making reweighting effective and allowing one simulation to deliver all PES-dependent results.
Second, the computational cost of PES evaluations can differ dramatically. In particular, when one PES provides analytic gradients while others require numerical differentiation, force evaluations for the latter may be orders of magnitude more expensive.
Propagating PIHMC on the low-cost reference PES while reweighting to the high-cost PESs can therefore reduce the overall cost by a comparable factor, with no change to the sampling procedure.
When the ensemble overlap is sufficient, the reweighted target-PES estimates can achieve statistical precision comparable to those from direct sampling, despite being obtained essentially for free once the reference trajectory is available.
The effectiveness of this strategy will be demonstrated  in \Sec{sec4-water-dimer} for the water dimer.

Furthermore, the same strategy may also prove useful in the development of PESs.
In practice, the construction of a PES often gives rise to a family of closely related variants for the same molecular system.
These variants may arise, for example, from different training sets\cite{Houston-JCP-2020-24107,Jing-JPCL-2025-10923}, various fitting ans\"{a}tze\cite{Houston-JCP-2020-24107,Qu-PCCP-2021-7758,Zhu-JCTC-2023-3551,Qu-MP-2024-2262058,Jing-JPCL-2025-10923}, the inclusion of small corrections\cite{Dral-JCP-2020-204110,Nandi-JACS-2023-9655}, or different members of a committee model.\cite{Schran-JCP-2020-104105}
Reweighting from a common reference ensemble then provides an efficient way to assess how such controlled changes in the surface affect tunneling observables, without repeating the full PIHMC sampling for every variant.
In this way, tunneling splittings can serve not only as benchmarks for a given PES, but also as sensitive diagnostics of which aspects of the surface are most important for quantitative accuracy.

\subsection{Post-processing of simulation data}\label{sec:post}
\subsubsection{Multistate Bennett acceptance ratio}\label{sec:mbar}

As discussed in \Sec{sec:windowing}, we partition the thermodynamic paths into overlapping windows and construct a local EBP for each window.
Using the PIHMC sampling scheme described above, we generate independent configurations from the corresponding window ensembles.
After sampling, all free-energy differences and prefactor averages required for the symmetrized partition functions are obtained via the Multistate Bennett acceptance ratio (MBAR) method.\cite{Shirts-JCP-2008-124105}
As MBAR is a well-established method, we present only a brief outline here and give a short summary in Appendix for completeness.

In MBAR, the free energies of all window ensembles are obtained by solving a set of self-consistent equations that optimally combines data from overlapping windows.
Once these free energies are known, the resulting MBAR weights yield statistically optimal estimates of ensemble averages for any target ring-polymer ensemble distribution whose reduced potential on the pooled samples is available.
In our implementation, this post-processing requires only the end-point bead coordinates of each sampled configuration.
If reweighting across PESs (described in \Sec{sec:reweight-pes}) is employed, it additionally requires the bead-summed physical potentials on the reference PES and each target PES for every sampled configuration.

\subsubsection{Error analysis}\label{sec:error}
To obtain tunneling splittings with the PIHMC-EBP framework, one must approach the limits $T\rightarrow0$ and $N\rightarrow\infty$.
In practice, simulations are performed at finite $T$ and finite $N$, and convergence is assessed by verifying that further lowering the temperature and increasing the number of beads does not change the results within statistical uncertainty.
Once this criterion is met, the resulting tunneling splittings are numerically exact for the selected PES.
Any remaining discrepancy with experiment is therefore attributable to the PES itself (or, more generally, to the underlying Born--Oppenheimer approximation).
For the bead-number convergence, one may additionally employ extrapolation with respect to $1/N$ to estimate the $N\to\infty$ limit.

In addition to these convergence considerations, the PIHMC-EBP simulation is subject to statistical errors due to finite sampling.
We estimate these uncertainties using a circular-block bootstrap\cite{Lahiri-ResamplingMethodsDependent-2003} that explicitly accounts for temporal correlations along the trajectories.
For each PIHMC trajectory, all sampled ring-polymer configurations are arranged in their natural time order, with the sequence closed cyclically.
For a trajectory containing $N_\mathrm{samp}$ in total, a block size of $N_{\mathrm{samp}}/500$ is chosen, and blocks of consecutive configurations are drawn with random starting points on the cycle and concatenated until a resampled dataset of the original length is obtained.
This choice of block size is typically much larger than the integrated autocorrelation time, while still allowing sufficient diversity in the resampled dataset.

In practice, a single calculation may consist of multiple independent PIHMC trajectories, for example from trivial parallelization within a window or from separate runs across different windows, and the complete dataset is formed by pooling the samples from all such trajectories.
In each bootstrap replicate, we resample each trajectory independently using the same circular-block procedure and then combine the resulting resampled trajectories to obtain a fully resampled dataset with the same number of independent trajectories as the original one.
Each resampled dataset is then analyzed with the same MBAR procedure as the original data.
Repeating this resampling and analysis many times yields an empirical distribution for every quantity of interest, from which standard deviations are taken as error estimates.
This approach preserves the intrinsic correlation structure of the original samples on the block scale without assuming the samples are uncorrelated, and is straightforward to implement for all observables derived from PIHMC-EBP, thus providing robust uncertainty estimates for the reported tunneling splittings and associated quantities.%

\subsection{Brief summary of the method}\label{sec:workflow}
The basic workflow of the PIHMC-EBP approach for computing tunneling splittings with rigorous symmetry and rotational projections can be summarized as follows.
\begin{enumerate}
    \item Identify the relevant symmetry operations and symmetrized partition functions $Z_P^{(J)}$.
    \item Construct the associated ring-polymer end-state potentials $\widetilde{U}_{P}(\bm{q})$, choose thermodynamic paths of interest, and pre-scan free-energy profiles using RPI estimates. The results are then used to construct an adaptive set of grid points ${\lambda_l}$ and to provide initial bias parameters ${\varepsilon_l}$ for the bridging potentials.
    \item Partition each thermodynamic path into multiple overlapping windows of moderate length and construct a local EBP in each window using the corresponding subset of grid points and biases.
Run short PIHMC simulations in all windows to obtain preliminary free-energy profiles and to assess overlap, using MBAR as a post-processing step.
    \item Refine the bias parameters of the bridging potentials based on these estimates and perform PIHMC simulations of sufficient length. During sampling, we store the endpoint bead coordinates $(\mathbf{q}^{(1)},\mathbf{q}^{(N)})$ for post-processing. If reweighting across PESs is employed, we additionally evaluate and store the bead-summed physical potential on every PES involved.
    \item Analyze simulation data using MBAR combined with circular-block bootstrapping to obtain tunneling splittings and other observables together with their statistical uncertainties.
\end{enumerate}

\section{RESULTS AND DISCUSSION\label{sec3}}

In this section, we apply PIHMC-EBP to compute ground-state tunneling splittings for malonaldehyde, the HCl dimer, and the water dimer.
Although the formulation is applicable to both rotational ground and rotationally excited manifolds, we focus here on the former.
For applications to tunneling splittings in rotationally excited states, see \rcite{Zupan-ExactTunnelingSplittings-2026}.
Numerically exact splittings within statistical uncertainty are obtained in the joint limits $T\rightarrow0$ and $N\rightarrow\infty$.
Accordingly, for each system we either choose a single temperature that is expected, based on existing benchmarks, to be sufficiently low, or we lower $T$ until the splittings converge.
At each selected temperature, we start from a moderate bead number $N$ and increase it until convergence within statistical uncertainty is achieved.

All PIHMC simulations employ a multiple-time-step velocity--Verlet integrator.
The outer time step $\Delta t$ is chosen to yield a propagation acceptance probability of $0.85-0.95$, while the ratio $\Delta t/\delta t$ is fixed at 8.
After each propagation update we record the endpoint bead coordinates required for the Eckart-spring estimators.
When reweighting is used to evaluate multiple PESs, we additionally store the physical-potential contributions to the ring-polymer energy on each PES.
For each simulation, 96 statistically independent trajectories are propagated in parallel and combined in the final analysis.

For each $(T,N)$ simulation, the EBP grid and initial biases are obtained from RPI calculations as described in \Sec{sec2}.
For large $N$, where the Hessian diagonalization required by RPI is expensive,  we reuse the grid and biases obtained at smaller $N$.
The grid is partitioned into several overlapping windows.
For each window, we first initialize from the RPI-optimized ring-polymer configuration at a representative $\lambda$ within the window, and perform a very short 100\,fs PIMD run (without a Metropolis accept–reject step) to relax the stiff optimized geometry, which is effective in stabilizing the acceptance ratio in the subsequent PIHMC propagation.
Starting from the relaxed configuration, we then carry out a short PIHMC run of 6\,ps, discarding the first 1\,ps and collecting data over the remaining 5\,ps to refine the biases in the EBP construction.
In this preliminary stage, the propagation length $L$ is fixed at 30 steps per update.

Using the refined biases from these short runs, we perform the long PIHMC simulations, initializing from the final configurations of preliminary trajectories and equilibrating for 5\,ps before data collection.
The propagation length $L$ is then chosen based on the autocorrelation analysis from the preliminary stage so that correlations decay rapidly over a small number of consecutive samples while the sampling remains sufficiently dense.
In both the preliminary short run and the long run, each propagation update is followed by five nonlocal permutation updates and $NL/20$ inter-bead rotation updates.

All MBAR analyses are performed using an in-house wrapper that calls the pymbar package\cite{PyMBAR403} to solve the self-consistent equations and evaluate averages in target distributions.
Statistical uncertainties are obtained from 500 circular-block bootstrap resamplings.
Additional system-specific simulation parameters are reported for  each molecular system considered below.
The raw grid-point data used in constructing the EBPs for all systems are provided in the Supplementary Material.

\subsection{Malonaldehyde}

We first apply the PIHMC-EBP scheme to both malonaldehyde and its deuterated isotopologue.
Malonaldehyde is a prototypical double-well system in which intramolecular proton transfer gives rise to a well-resolved ground-state tunneling splitting that has been accurately measured experimentally.\cite{Baughcum-JACS-1984-2260,Turner-JACS-1984-2265,Firth-JCP-1991-1812,Baba-JCP-1999-4131}
From a theoretical standpoint, malonaldehyde contains nine atoms, making fully quantum treatments challenging while still permitting high-quality analytic PESs fitted to benchmark electronic-structure data, and has therefore served as a standard test for theoretical methods.

The most accurate PES for malonaldehyde currently available is the one reported by Mizukami \textit{et al}.\cite{Mizukami-JCP-2014-144310}
Based on this PES, tunneling splittings have been computed using a range of approaches, including DMC\cite{Mizukami-JCP-2014-144310}, RPI with and without perturbative corrections\cite{Lawrence-JCP-2023-14111}, variational calculations on Smolyak grids\cite{Lauvergnat-C-2023-202300501}, and different implementations of PIMD.\cite{Matyus-JCP-2016-114108,Vaillant-JCP-2018-234102,Baumann-MP-2025-2474202}
For the normal isotopologue, the numerically exact splitting with a relative error smaller than 1\% was only recently obtained using an Eckart-spring PIMD formulation\cite{Baumann-MP-2025-2474202}, while that for the deuterated species has not yet been published.

Here, we employ this PES and use PIHMC-EBP to compute the tunneling splittings of both normal and deuterated malonaldehyde.
This allows a direct comparison with existing approaches, particularly PIMD-based schemes, and demonstrates the advantages of the present methodology.
At the same time, our calculations also provide new high-accuracy reference values for the tunneling splittings of both isotopologues.

\begin{figure}[htbp]
    \centering
    \includegraphics[width=0.98\columnwidth]{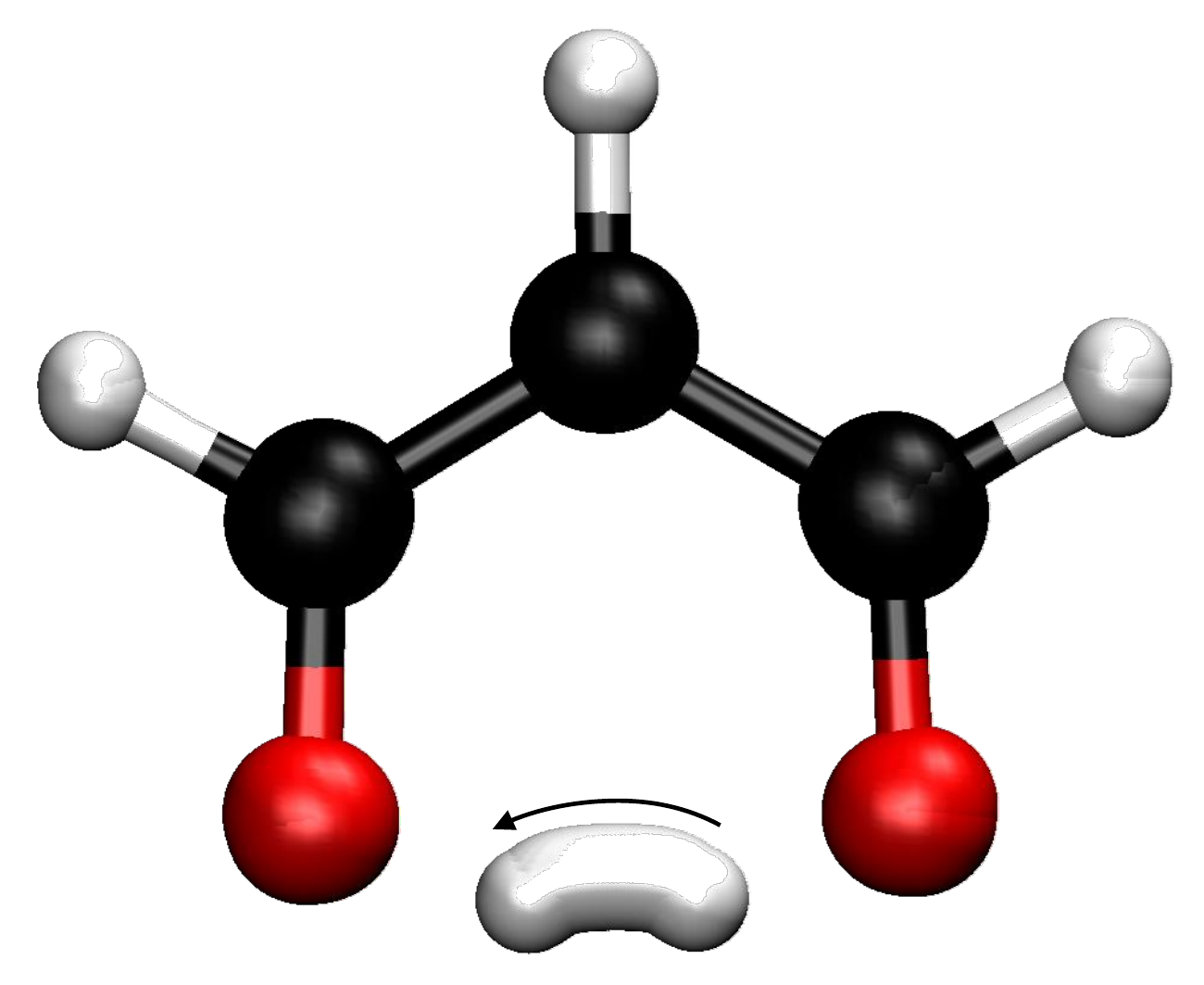}
    \caption{Instanton configuration for malonaldehyde associated with the \symop{P} end state at $\lambda=1$. The arrow indicates the direction of proton  transfer.}
    \label{fig:mht-h-kink}
\end{figure}

For malonaldehyde, the symmetry associated with the tunneling motion is described by a two-element group generated by the identity $\symop{E}$ and the permutation $\symop{P}=(15)(24)(79)$, where the atom labeling $612738495$ follows \rcite{Baughcum-JACS-1981-6296}.
This group is isomorphic to $S_2$, and its character table is given in \Tab{tab:s2-ct}.
In the EBP construction, we use a family of intermediate potentials $U_\mathrm{m}(\bm{q};\lambda)$ connecting the two target ring-polymer potentials $\widetilde{U}_{\symop{E}}(\bm{q})$ and $\widetilde{U}_{\symop{P}}(\bm{q})$, corresponding to $\lambda=0$ and $\lambda=1$, respectively.
\Fig{fig:mht-h-kink} shows the instanton path on the $\lambda=1$ endpoint for malonaldehyde, while the corresponding instanton on the intermediate potential at $\lambda=0.5$ is shown in Fig.\,S1 of the Supplementary Material.

\begin{table}[htbp]
\caption{Character table of the symmetric group $S_2$.}
\label{tab:s2-ct}
{\renewcommand{\arraystretch}{1.5}
\begin{ruledtabular}
\begin{tabular}{lcc}
$\Gamma$ & $\phantom{-}\symop{E}$ & $\symop{P}$ \\
\hline
$\irrep{A}$ & $\phantom{-}1$ & $\phantom{-}1$ \\
$\irrep{B}$ & $\phantom{-}$1 & $-1$ \\
\end{tabular}
\end{ruledtabular}
}
\end{table}

In PIHMC simulations for malonaldehyde, we set $T=50\,\mathrm{K}$, since a previous six-state-model analysis\cite{Baumann-MP-2025-2474202} indicates that contamination from vibrationally excited states at this temperature contributes only $\sim0.01\,\mathrm{cm}^{-1}$ to the splitting, well below our statistical uncertainty.
We consider $N=256$, 512, and 768, with propagation length $L=30$ for $N=256$ and $L=60$ for $N=512$ and 768.
For deuterated malonaldehyde we set $T=25\,\mathrm{K}$ and $N=512$, 768, and 1024, with $L=30$ at $N=512$ and $L=60$ otherwise, as the frequency reduction upon deuteration is modest and this temperature should be sufficient to suppress vibrational contributions.
In all cases, the EBP grid is partitioned into 5 overlapping windows of equal size, with the corresponding grid points and bias parameters reported in the Supplementary Material, and each long PIHMC run consists of 5\,ps equilibration followed by 800\,ps of data collection.

The PIHMC-EBP results for malonaldehyde are summarized in the last column of \Tab{tab:mhth}.
The tunneling splittings obtained at $N=256$, 512, and 768 are $20.87(7)$, $21.27(9)$, and $21.31(9)\,\mathrm{cm^{-1}}$, respectively.
The change from $N=512$ to 768 is not statistically significant, indicating that the splitting is converged within uncertainty.
We therefore report the $N=768$ estimate $21.31(9)\,\mathrm{cm^{-1}}$ as the final, numerically exact tunneling splitting of malonaldehyde on this PES.

\begin{table}[htbp]
\caption{Tunneling splitting, $\Delta$, of malonaldehyde calculated with PIHMC-EBP and PIMD-TI at $T=50\,\mathrm{K}$ and different bead numbers $N$. Values in parentheses denote one standard error in the last digit quoted. In the bottom part of the table, the converged PIHMC-EBP result (taken from $N=768$, highlighted in bold in the table) is compared with the PIMD-TI and experimentally measured (Expt.) values.}
\label{tab:mhth}
{\renewcommand{\arraystretch}{1.5}
\begin{ruledtabular}
\begin{tabular*}{\columnwidth}{@{\extracolsep{\fill}}lccccc}
 & & & \multicolumn{3}{c}{$\Delta\,(\mathrm{cm}^{-1})$} \\
\cline{4-6}
            &         &         & \multicolumn{2}{c}{PIMD\cite{Baumann-MP-2025-2474202}\footnotemark[1]} &     \multirow{2}{*}{PIHMC\footnotemark[4]}    \\
\cline{4-5}
$T$\,(K)    &   $N$   &  $\Delta t\,(\mathrm{fs})$  &   $n_\xi=10$   &  $n_\xi=15$   &   \\
\hline
50          &  256    &  0.25   &  $21.2(2)$  &  $21.1(2)$  & $20.87(7)$  \\
            &         &  0.125  &  $20.6(2)$  &  $20.9(2)$  & $-$         \\
            &  512    &  0.25   &  $21.1(3)$  &  $22.0(2)$  & $-$         \\
            &         &  0.125  &  $21.2(3)$  &  $21.6(2)$  & $21.27(9)$  \\
            &  768    &  0.1    &     $-$     &    $-$      & $\B{21.31(9)}$  \\
\hline
Method                       & \multicolumn{2}{c}{Expt.\cite{Turner-JACS-1984-2265}}  & ElVibRot\cite{Lauvergnat-C-2023-202300501}\footnotemark[2]& PIMD\cite{Baumann-MP-2025-2474202}\footnotemark[3] & PIHMC \\
$\Delta\,(\mathrm{cm}^{-1})$ &  \multicolumn{2}{c}{$21.583$}    &  $21.7(3)$ &  $21.1(1)$    &  $21.31(9)$       \\
\end{tabular*}
\end{ruledtabular}
\footnotetext[1]{$n_\xi$ is the number of quadrature points in PIMD-TI simulations.}
\footnotetext[2]{Uncertainty in parentheses is an estimate of the basis-set incompleteness error.}
\footnotetext[3]{Obtained from a Bayesian analysis of many independent PIMD-TI calculations with varied parameters.}%
\footnotetext[4]{This work.}
}
\end{table}

The last row of \Tab{tab:mhth} compares our result with previous numerically exact theoretical benchmarks as well as the experimentally measured value.
The ElVibRot value, $21.7(3)\,\mathrm{cm^{-1}}$, was obtained from a Smolyak-grid variational calculation\cite{Lauvergnat-C-2023-202300501}, and the quoted uncertainty reflects an estimate of the residual basis-set incompleteness error.
Its lower bound, $21.4\,\mathrm{cm^{-1}}$, overlaps with our PIHMC-EBP result, $21.31(9)\,\mathrm{cm^{-1}}$, at the $1\sigma$ level, so the two calculations are statistically consistent.
At the same time, our uncertainty is about three times smaller, providing a substantially tighter reference value for this PES.

The PIMD-TI benchmark\cite{Baumann-MP-2025-2474202}, $21.1(1)\,\mathrm{cm^{-1}}$, differs from our result by $0.21\,\mathrm{cm^{-1}}$.
Treating the quoted uncertainties as independent standard errors, the corresponding combined uncertainty is $\sqrt{0.09^2+0.10^2}=0.134\,\mathrm{cm^{-1}}$, so the deviation amounts to approximately $1.6\sigma$.
This level of discrepancy is not statistically significant, although it suggests a modest tension.
We note that the PIMD-TI value was obtained from a Bayesian analysis based on many independent calculations performed with varied numerical settings ($T$, $N$, $\Delta t$, and the number of quadrature points $n_\xi$), thereby accounting for possible residual systematic biases of individual calculation.
By contrast, our PIHMC-EBP estimate is taken from the simulation at $N=768$, after verifying that the residual errors associated with finite $T$ and $N$ are smaller than the statistical uncertainty, and therefore it does not rely on additional assumptions about the distribution of systematic biases across runs.
On this basis, we regard $21.31(9)\,\mathrm{cm^{-1}}$ as the most precise estimate currently available for the malonaldehyde ground-state tunneling splitting on this PES.
Compared with the experimental value\cite{Turner-JACS-1984-2265} of $21.583\,\mathrm{cm^{-1}}$, our high-precision result provides a stringent benchmark for the intrinsic error of this PES, which underestimates the splitting by $0.27(9)\,\mathrm{cm^{-1}}$ or about $1.3\%$.

To facilitate a direct numerical comparison between PIHMC-EBP and PIMD-TI, \Tab{tab:mhth} also lists a set of PIMD-TI results obtained at $T=50\,\mathrm{K}$ using different parameter choices.
In all PIMD-TI calculations\cite{Baumann-MP-2025-2474202}, the sampling length is 25\,ps per TI point and 96 statistically independent trajectories are propagated in parallel.
As a representative example, the setting $(N,\Delta t,n_\xi)=(512, 0.125\,\mathrm{fs}, 15)$ corresponds to a total propagation time of $96\times15\times0.025=36\,\mathrm{ns}$ for sampling, or around $1.5\times10^{11}$ evaluations of the physical potential and its gradient.
Accounting for all four PIMD simulations at $N=512$ in \Tab{tab:mhth}, the total cost is approximately $3.7\times10^{11}$ physical potential and gradient evaluations.
The resulting splitting, $21.6(2)\,\mathrm{cm}^{-1}$, has a statistical uncertainty of $0.2\,\mathrm{cm}^{-1}$.

By comparison, the PIHMC-EBP calculation at $N=512$ consists of 96 independent trajectories of 800\,ps in each of 5 windows, corresponding to a total propagation time of $96\times5\times0.8=384\,\mathrm{ns}$ for sampling, or approximately $1.6\times10^{12}$ calculations of the physical potential and its gradient.
This yields $21.27(9)\,\mathrm{cm}^{-1}$, with a statistical uncertainty of $0.09\,\mathrm{cm}^{-1}$.
Assuming purely statistical error reduction proportional to the inverse square root of sampling length, a naive comparison would suggest that PIHMC-EBP is slightly less expensive for the same accuracy target (about 0.88 times the cost of PIMD-TI).

However, this comparison is perhaps too conservative, as the individual PIMD-TI data points summarized in \Tab{tab:mhth} from \rcite{Baumann-MP-2025-2474202} were not intended as a systematic convergence study of $\Delta t$ and $n_\xi$ at the $0.2\,\mathrm{cm}^{-1}$ uncertainty level.
For $n_\xi$, this is illustrated most clearly at $N=512$ and $\Delta t=0.25\,\mathrm{fs}$, where increasing $n_\xi$ from 10 to 15 shifts the splitting from $21.1(3)$ to $22.0(2)\,\mathrm{cm}^{-1}$.
The corresponding change of $0.9\,\mathrm{cm}^{-1}$ amounts to $2.5\sigma$ when the reported uncertainties are combined, indicating a statistically significant dependence of the results on $n_\xi$ within this range.
A similarly pronounced sensitivity is observed for $\Delta t$ at $N=256$ and $n_\xi=10$, where reducing $\Delta t$ from 0.25 to $0.125\,\mathrm{fs}$ shifts the splitting from $21.2(2)$ to $20.6(2)\,\mathrm{cm}^{-1}$, corresponding to $2.1\sigma$ variation.
Given this clear $\Delta t$ dependence observed at $N=256$, one may expect more stringent $\Delta t$ requirements at larger $N$, for which the ring-polymer springs are stiffer.
To establish convergence more rigorously, at least one additional refinement in each parameter would be required.
\footnote{other cross-checks of  the PIMD-TI data in \Tab{tab:mhth} appear mutually consistent, for example when comparing the two $n_\xi$ values at $N=512$ and $\Delta t=0.125\,\mathrm{fs}$, or when comparing the two $\Delta t$ values at $N=512$ and $n_\xi=15$.
This observation, which seems to contradict the discussion above, actually suggests strong cancellation between residual errors in $\Delta t$ and $n_\xi$ for certain parameter combinations.
Therefore, the seemingly statistical consistency between selected entries in \Tab{tab:mhth} should not be interpreted as evidence that both parameters converge.}

These additional refinement calculations would be even more expensive than those already reported.
As a simple lower-bound estimate based on the existing data points, consider the minimal incremental refinement set $\Delta t=0.25, 0.125, 0.0625\,\mathrm{fs}$ together with $n_\xi=10, 15, 20$ for $N=512$, which requires nine PIMD-TI calculations in total.
If one assumes that the same simulation setup yields a comparable statistical uncertainty for each additional calculation, then roughly $1.6\times10^{12}$ evaluations of the physical potential and the gradient would be required to reach the $0.2\,\mathrm{cm}^{-1}$ uncertainty level, which is already about 5 times the cost required by PIHMC-EBP for the same statistical precision.
Furthermore, this lower-bound estimate is based on the minimal refinement set required to reach $0.2\,\mathrm{cm}^{-1}$ statistical uncertainty.
Achieving $0.09\,\mathrm{cm}^{-1}$ uncertainty would probably require additional refinements, potentially pushing the total PIMD-TI cost to more than one order of magnitude above PIHMC-EBP.

More importantly, the human effort required differs substantially.
As is already apparent from the discussion above, since PIMD-TI involves the convergence checks of $\Delta t$ and $n_\xi$, obtaining a high-precision result typically requires a nontrivial statistical analysis of multiple partially converged calculations and successive runs with increasingly expensive settings.
Many of these runs serve primarily to diagnose residual bias and are not directly used in the final estimate, which is inherently wasteful.
PIHMC-EBP avoids this overhead, as the accuracy for a chosen $(T,N)$ can be improved without additional convergence studies by directly increasing the sampling length, thereby substantially simplifying the workflow and dramatically reducing manual effort.

\begin{table}[htbp]
\caption{Tunneling splitting $\Delta$ of deuterated malonaldehyde calculated with PIHMC-EBP at $T=25\,\mathrm{K}$ and different bead numbers $N$. Values in parentheses denote one standard error in the last digit quoted. In the bottom part of the table, the converged PIHMC-EBP result (taken from $N=1024$, highlighted in bold in the table) is compared with a Smolyak-grid-based wavefunction approach (ElVibRot) and the experimentally measured (Expt.) values.}
\label{tab:mhtd}
{\renewcommand{\arraystretch}{1.5}
\begin{ruledtabular}
\begin{tabular*}{\columnwidth}{@{\extracolsep{\fill}}lccc}
$T$\,(K)    &   $N$   &   $\Delta t\,(\mathrm{fs})$  &    $\Delta\,(\mathrm{cm}^{-1})$\\
\hline
$25$      & $512$     &   0.25     &   $2.80(1)$ \\
          & $768$     &   0.15     &   $2.83(1)$ \\
          & $1024$    &   0.125    &   $\B{2.84(1)}$ \\
\hline
Method                       & Expt.\cite{Baughcum-JACS-1984-2260} & ElVibRot\cite{Lauvergnat-C-2023-202300501} & PIHMC \\
$\Delta\,(\mathrm{cm}^{-1})$ &   $2.915(4)$   &      $2.9(1)$\footnotemark[1]    &     $2.84(1)$ \\
\end{tabular*}
\end{ruledtabular}
\footnotetext[1]{Uncertainty in parentheses is an estimate of the basis-set incompleteness error.}
}
\end{table}

\Tab{tab:mhtd} summarizes the PIHMC-EBP results for deuterated malonaldehyde and compares them with the experimental value and the ElVibRot benchmark.
The splittings obtained at $N=512$, 768, and 1024 are $2.80(1)$, $2.83(1)$, and $2.84(1)\,\mathrm{cm^{-1}}$, respectively, indicating convergence with respect to $N$ within statistical uncertainty.
We therefore report the numerically exact tunneling splitting on this PES as $2.84(1)\,\mathrm{cm^{-1}}$, taken from the $N=1024$ calculation.
This value is consistent with the ElVibRot result of $2.9(1)\,\mathrm{cm^{-1}}$ while reducing the uncertainty by roughly an order of magnitude.
Therefore, to our knowledge, it is the most precise theoretical estimate currently available for the deuterated isotopologue on this PES.
Comparison with the experimental value\cite{Baughcum-JACS-1984-2260} $2.915(4)\,\mathrm{cm^{-1}}$ indicates a relative  deviation of about $2.6\%$ for the deuterated splitting due to the intrinsic error of the PES.

\subsection{HCl dimer}

We next apply the PIHMC-EBP approach to the HCl dimer.
As one of the simplest hydrogen halide clusters, this system provides a prototypical platform for investigating hydrogen-bond interactions and nuclear quantum effects.
Very recently, Shen and co-workers\cite{Shen-JCP-2025-154307} constructed a high-accuracy many-body machine-learned PES for HCl clusters and, using PIMD-TI with the Eckart spring, computed the numerically exact ground-state tunneling splitting of the dimer.
However, severe positive and negative cancellations in the TI lead to substantial statistical uncertainties, even with very long simulations.
This system is therefore particularly suitable for assessing the advantage of the present approach in this regard.

\begin{figure}[htbp]
    \centering
    \includegraphics[width=0.98\columnwidth]{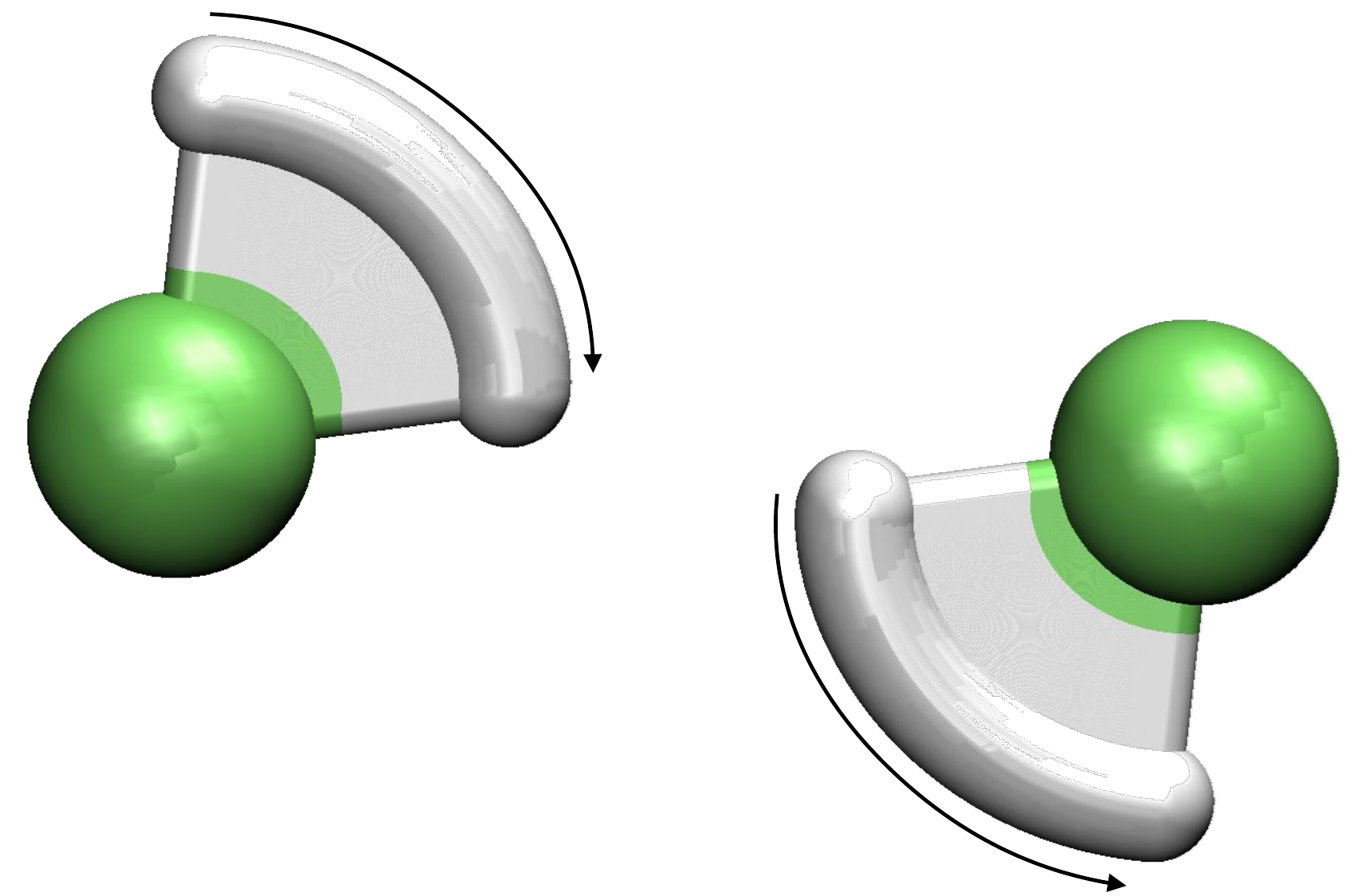}
    \caption{Instanton configuration for the HCl dimer associated with the \symop{P} end state at $\lambda=1$. The arrow indicates the direction of hydrogen motion.}
    \label{fig:hcl2-kink}
\end{figure}

For the HCl dimer, the tunneling motion corresponds to interchange of the donor and acceptor monomers, which is described by the permutation operation
$\symop{P}=(\mathrm{H}_1\,\mathrm{H}_2)(\mathrm{Cl}_1\,\mathrm{Cl}_2)$.
Similarly to malonaldehyde, this together with the identity $\symop{E}$ generates a two-element group isomorphic to $S_2$, with the character table given in \Tab{tab:s2-ct}.
The EBP is thus constructed along a thermodynamic path that interpolates between the target ring-polymer potentials associated with $\symop{E}$ and $\symop{P}$, corresponding to $\lambda=0$ and 1, respectively.
\Fig{fig:hcl2-kink} shows the instanton configuration at $\lambda=1$ associated with $\symop{P}$, while the corresponding instanton for the intermediate bridging potential at $\lambda=0.5$ is shown in Fig.~S2 of the Supplementary Material.

In the PIHMC-EBP calculations we consider $T=20$ and $10\,\mathrm{K}$ with bead numbers $N=128$, 256, and 512, to facilitate comparison with the previous PIMD-TI result that was obtained at $10\,\mathrm{K}$ and $N=256$.
For each set of parameters $(T,N)$, the EBP grid is partitioned into 3 overlapping windows, with the grid points and bias parameters reported in the Supplementary Material.
Each long PIHMC run consists of $5\,\mathrm{ps}$ equilibration followed by $100\,\mathrm{ps}$ of data collection.
The propagation length is set to $L=30$ in all cases, except for $(T,N)=(10\,\mathrm{K},512)$ where $L=60$ is used.

\begin{table}[htbp]
\caption{Tunneling splitting $\Delta$ of the HCl dimer calculated with PIHMC-EBP at different temperatures $T$ and bead numbers $N$. Values in parentheses denote one standard error in the last quoted digit. In the bottom part of the table, the converged PIHMC-EBP result (taken from $T=10\,\mathrm{K}$ and $N=512$, highlighted in bold in the table) is compared with the PIMD-TI and experimentally measured (Expt.) values.}
\label{tab:hcl2}
{\renewcommand{\arraystretch}{1.5}
\begin{ruledtabular}
\begin{tabular*}{\columnwidth}{@{\extracolsep{\fill}}lccc}
$T$\,(K)    &   $N$   &   $\Delta t\,(\mathrm{fs})$ & $\Delta\,(\mathrm{cm}^{-1})$ \\
\hline
$20$      & $128$ &   0.5  &  $13.4(2)$ \\
          & $256$ &   0.4  &  $13.3(2)$ \\
$10$      & $128$ &   0.5  &  $13.4(3)$ \\
          & $256$ &   0.4  &  $12.9(3)$ \\
          & $512$ &   0.3  &  $\B{13.1(3)}$ \\
\hline
Method & Expt.\cite{Schuder-JCP-1989-4418} & PIMD\cite{Shen-JCP-2025-154307} & PIHMC \\
$\Delta\,(\mathrm{cm}^{-1})$ & 15.46 & $13.9(9)$ & $13.1(3)$ \\
\end{tabular*}
\end{ruledtabular}
}
\end{table}

\Tab{tab:hcl2} summarizes the PIHMC-EBP splittings obtained at different $T$ and $N$ and also lists the corresponding outer time steps $\Delta t$.
The value $13.1(3)\,\mathrm{cm}^{-1}$ from $T=10\,\mathrm{K}$ and $N=512$  is statistically consistent with the results at smaller $N$ and at $T=20\,\mathrm{K}$, indicating that the splitting is converged within uncertainty.
We therefore take $13.1(3)\,\mathrm{cm}^{-1}$ as the numerically exact ground-state tunneling splitting of the HCl dimer on this PES.
This is consistent with the previous PIMD-TI estimate of $13.9(9)\,\mathrm{cm}^{-1}$, but reduces the uncertainty by roughly a factor of three.
Comparison with the experimental value\cite{Schuder-JCP-1989-4418}  $15.46\,\mathrm{cm}^{-1}$ shows that the relative error of this state-of-the-art PES for the tunneling splitting is about 15\%, rather than below 10\%, as previously inferred from the earlier, lower-precision PIMD-TI result.
We hope that the present high-precision benchmark will help guide further development of spectroscopic-quality PESs for the HCl dimer and other H-bonded systems.

We now compare the computational cost of the two approaches.
According to \rcite{Shen-JCP-2025-154307}, the PIMD-TI calculation was performed at $T=10\,\mathrm{K}$ and $N=256$ using a single trajectory of 13\,ns with $\Delta t=0.1\,\mathrm{fs}$ and $n_\xi=23$, corresponding to approximately $7.7\times10^{11}$ evaluations of the  physical potential and its gradient and yielding a statistical uncertainty of about $0.9\,\mathrm{cm}^{-1}$.
With the same $T$ and $N$ values, our PIHMC-EBP calculations use 96 independent trajectories of 
100\,ps in each of 3 windows with $\Delta t=0.4\,\mathrm{fs}$, corresponding to $1.8\times10^{10}$ physical-potential and gradient evaluations and giving a statistical uncertainty of $0.3\,\mathrm{cm}^{-1}$.
Since statistical errors scale inversely with the square root of the number of samples, these figures imply that, for a single parameter setting, PIMD-TI is roughly 380 times more expensive than PIHMC-EBP at comparable accuracy.
Once the additional PIMD-TI calculations required to assess convergence with respect to $\Delta t$ and $n_\xi$ are included, the overall cost advantage of PIHMC-EBP can readily approach three orders of magnitude.
If one additionally targets the same $0.3\,\mathrm{cm}^{-1}$ uncertainty with PIMD-TI, the convergence assessment would become more demanding and the required computational effort would increase further.

\subsection{Water dimer}\label{sec4-water-dimer}

\begin{table*}[htbp]
\caption{Character table of the permutation--inversion group $G_{16}$ for the water dimer (isomorphic to $D_\mathrm{4h}$).
The notation for symmetry operations, atom labeling, and monomer definitions follows \rcite{Dyke-JCP-1977-492}.}
\label{tab:water-dimer-ct}
\renewcommand{\arraystretch}{1.5}
\begin{ruledtabular}
\begin{tabular*}{\textwidth}{@{\extracolsep{\fill}}lcccccccccc}
$\Gamma$ &
$\symop{E}$ &
\begin{tabular}{@{}c@{}}$(12)$\\$(34)$\end{tabular} &
\begin{tabular}{@{}c@{}}$(\symop{ab})(13)(24)$\\$(\symop{ab})(14)(23)$\end{tabular} &
\begin{tabular}{@{}c@{}}$(\symop{ab})(1324)$\\$(\symop{ab})(1423)$\end{tabular} &
$(12)(34)$ &
$\phantom{-}\symop{E}^{*}$ &
\begin{tabular}{@{}c@{}}$(12)^{*}$\\$(34)^{*}$\end{tabular} &
\begin{tabular}{@{}c@{}}$(\symop{ab})(13)(24)^{*}$\\$(\symop{ab})(14)(23)^{*}$\end{tabular} &
\begin{tabular}{@{}c@{}}$(\symop{ab})(1324)^{*}$\\$(\symop{ab})(1423)^{*}$\end{tabular} &
$(12)(34)^{*}$ \\
\hline
$\irrep{A}_{1}^{+}$ & $1$ &  $\phantom{-}1$ &  $\phantom{-}1$ &  $\phantom{-}1$ &  $\phantom{-}1$ &  $\phantom{-}1$ &  $\phantom{-}1$ &  $\phantom{-}1$ &  $\phantom{-}1$ &  $\phantom{-}1$ \\
$\irrep{A}_{2}^{+}$ & $1$ & $-1$ & $-1$ &  $\phantom{-}1$ &  $\phantom{-}1$ &  $\phantom{-}1$ & $-1$ & $-1$ &  $\phantom{-}1$ &  $\phantom{-}1$ \\
$\irrep{B}_{1}^{+}$ & $1$ &  $\phantom{-}1$ & $-1$ & $-1$ &  $\phantom{-}1$ &  $\phantom{-}1$ &  $\phantom{-}1$ & $-1$ & $-1$ &  $\phantom{-}1$ \\
$\irrep{B}_{2}^{+}$ & $1$ & $-1$ &  $\phantom{-}1$ & $-1$ &  $\phantom{-}1$ &  $\phantom{-}1$ & $-1$ &  $\phantom{-}1$ & $-1$ &  $\phantom{-}1$ \\
$\irrep{E}^{+}$     & $2$ &  $\phantom{-}0$ &  $\phantom{-}0$ &  $\phantom{-}0$ & $-2$ &  $\phantom{-}2$ &  $\phantom{-}0$ &  $\phantom{-}0$ &  $\phantom{-}0$ & $-2$ \\
$\irrep{A}_{1}^{-}$ & $1$ &  $\phantom{-}1$ &  $\phantom{-}1$ &  $\phantom{-}1$ &  $\phantom{-}1$ & $-1$ & $-1$ & $-1$ & $-1$ & $-1$ \\
$\irrep{A}_{2}^{-}$ & $1$ & $-1$ & $-1$ &  $\phantom{-}1$ &  $\phantom{-}1$ & $-1$ &  $\phantom{-}1$ &  $\phantom{-}1$ & $-1$ & $-1$ \\
$\irrep{B}_{1}^{-}$ & $1$ &  $\phantom{-}1$ & $-1$ & $-1$ &  $\phantom{-}1$ & $-1$ & $-1$ &  $\phantom{-}1$ &  $\phantom{-}1$ & $-1$ \\
$\irrep{B}_{2}^{-}$ & $1$ & $-1$ &  $\phantom{-}1$ & $-1$ &  $\phantom{-}1$ & $-1$ &  $\phantom{-}1$ & $-1$ &  $\phantom{-}1$ & $-1$ \\
$\irrep{E}^{-}$     & $2$ &  $\phantom{-}0$ &  $\phantom{-}0$ &  $\phantom{-}0$ & $-2$ & $-2$ &  $\phantom{-}0$ &  $\phantom{-}0$ &  $\phantom{-}0$ &  $\phantom{-}2$ \\
\end{tabular*}
\end{ruledtabular}
\end{table*}

\begin{figure*}[htbp]
    \centering
    \includegraphics[width=0.98\linewidth]{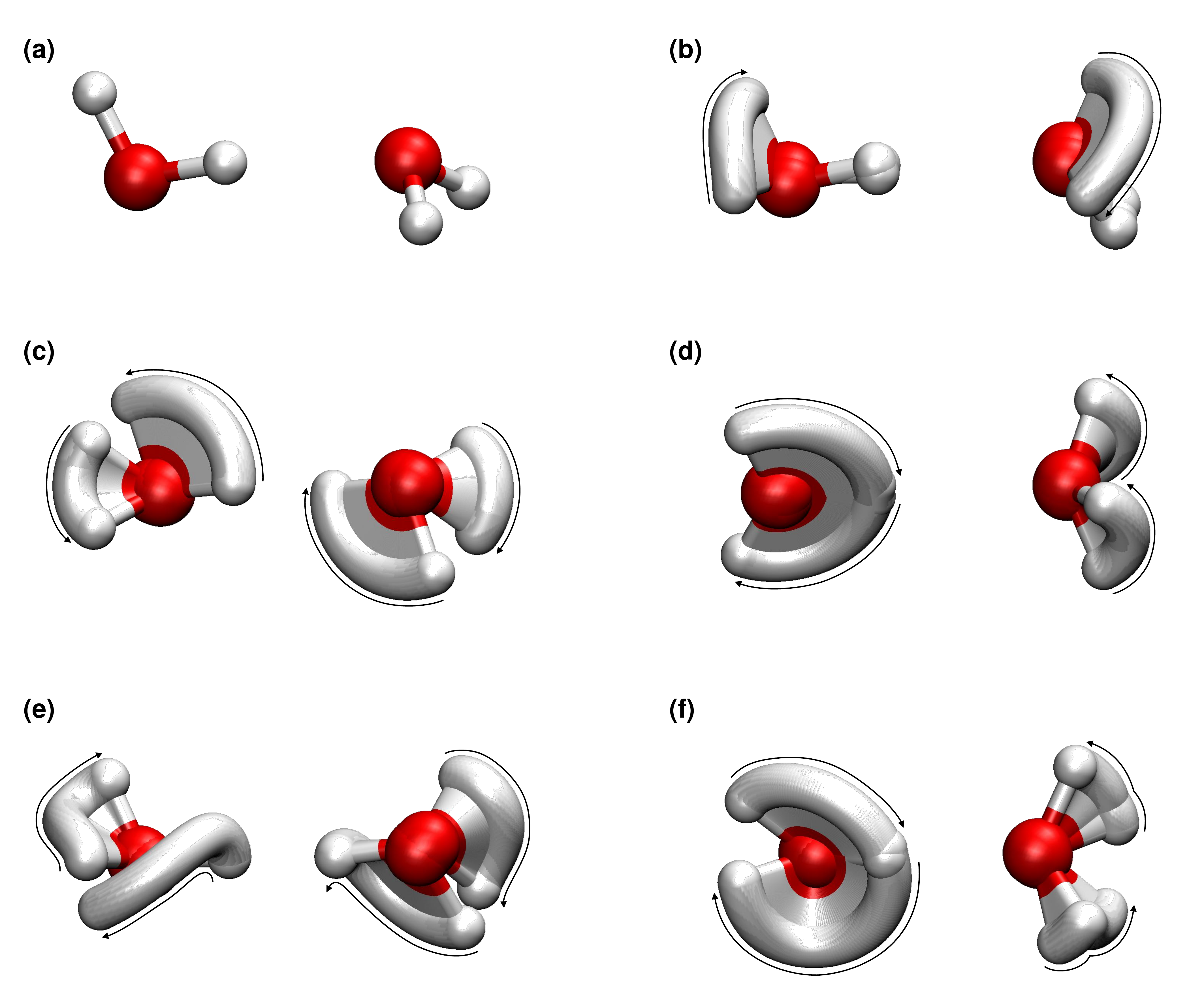}
    \caption{Minimum-action instanton configurations for the six EBP end states of the water dimer, with panels (a)--(f) corresponding to the composite symmetry operation sets 1--6 in \Tab{tab:sym-sets} in the same order. The tunneling motions are (a) identity, (b) acceptor tunneling, (c) geared interchange, (d) bifurcation tunneling, (e) anti-geared interchange, and (f) donor exchange. Arrows indicate the corresponding hydrogen rearrangements.}
    \label{fig:water-dimer-kink}
\end{figure*}

The final example is the water dimer.
As the simplest hydrogen-bonded water cluster, it serves as a stringent test of pairwise interaction models and a direct link between microscopic hydrogen bonding and accurately measured vibration-rotation-tunneling spectra.
Over the years, numerous high-level PESs have been developed for this system. %
Here, we consider three water-dimer PESs, the MB-pol dimer model\cite{Babin-JCTC-2013-5395}, CCpol-8sfIR[2012] with $f=1$ embedding\cite{Leforestier-JCP-2012-14305,Wang-JCP-2018-74108} (hereinafter referred to as CCpol-8sf), and the recently developed flex-CCpol2025 surface.\cite{Jing-JPCL-2025-10923}

This choice is motivated first by the availability of reliable benchmarks.
For MB-pol, previous results are available both from variational calculations within the adiabatic approximation\cite{Babin-JCTC-2013-5395} and from PIMD-TI.\cite{Vaillant-JCP-2018-234102}
The former provide an accurate, although not numerically exact, reference, whereas the latter suffer from severe contamination by rotationally excited states.
By contrast, CCpol-8sf and flex-CCpol2025 have numerically exact variational benchmarks.
Second, the three PESs differ markedly in computational cost.
MB-pol provides analytic gradients, while CCpol-8sf and flex-CCpol2025 require numerical differentiation, making force evaluations roughly two orders of magnitude more expensive.
This cost separation makes the system well suited to the reweighting strategy introduced in \Sec{sec:reweight-pes}, in which the PIHMC trajectories are propagated using MB-pol while reweighting is used to extract splittings on all three PESs.
Compared with direct simulations on the latter two surfaces, this trick reduces the overall computational cost by two orders of magnitude.

The feasible permutation--inversion symmetry of the water dimer, restricted to operations that interchange identical nuclei without bond breaking, is described by the $G_{16}$ group, which is isomorphic to $D_\mathrm{4h}$.
Following the labeling convention of \rcite{Dyke-JCP-1977-492}, the elements include permutations of the two equivalent protons within each monomer, interchange of the two monomers, and their combinations with inversion through the center of mass, giving 16 symmetry operations distributed over 10 conjugacy classes.
The corresponding character table is listed in \Tab{tab:water-dimer-ct}, which contains 10 irreps, namely 8 one-dimensional irreps \irrep{A_{1}^{\pm}}, \irrep{A_{2}^{\pm}}, \irrep{B_{1}^{\pm}}, \irrep{B_{2}^{\pm}}, and 2 two-dimensional irreps \irrep{E^{\pm}}.
In the present work, we focus on the $J=0$ rovibrational ground-state tunneling splittings, which form two triplets conventionally labeled the $1s$ and $2s$ manifolds and comprise the six symmetry species \irrep{A_{1}^{+}}, \irrep{B_{1}^{+}}, \irrep{E^{+}}, \irrep{A_{2}^{-}}, \irrep{B_{2}^{-}}, and \irrep{E^{-}}.

\begin{table}[htbp]
\caption{Composite sets of symmetry operations defining the end states used for the EBP construction, obtained by merging conjugacy classes in \Tab{tab:water-dimer-ct}. Conjugacy classes are numbered 1--10 in the same order as in \Tab{tab:water-dimer-ct}.}
\label{tab:sym-sets}
{\renewcommand{\arraystretch}{1.5}
\begin{ruledtabular}
\begin{tabular*}{\columnwidth}{@{\extracolsep{\fill}}ccl}
Set & Classes & Symmetry operations \\
\hline
\multirow{2}{*}{1} & 1 & $\symop{E}$ \\
                   & 7 & $(12)^{*}$,\ $(34)^{*}$ \\
\hline
\multirow{2}{*}{2} & 2 & $(12)$,\ $(34)$ \\
                   & 6 & $\symop{E}^{*}$ \\
\hline
\multirow{2}{*}{3} & 3 & $(\symop{ab})(13)(24)$,\ $(\symop{ab})(14)(23)$ \\
                   & 9 & $(\symop{ab})(1324)^{*}$,\ $(\symop{ab})(1423)^{*}$ \\
\hline
\multirow{2}{*}{4} & 4 & $(\symop{ab})(1324)$,\ $(\symop{ab})(1423)$ \\
                   & 8 & $(\symop{ab})(13)(24)^{*}$,\ $(\symop{ab})(14)(23)^{*}$ \\
\hline
5 & 5  & $(12)(34)$ \\
\hline
6 & 10 & $(12)(34)^{*}$ \\
\end{tabular*}
\end{ruledtabular}
}
\end{table}

\begin{table*}[htbp]
\caption{Energy-level splittings (cm$^{-1}$) of the water dimer on the CCpol-8sf\cite{Leforestier-JCP-2012-14305,Wang-JCP-2018-74108}, flex-CCpol2025\cite{Jing-JPCL-2025-10923}, and MB-pol\cite{Babin-JCTC-2013-5395} PESs from PIHMC-EBP simulations, wavefunction-based (WFN) calculations, and PIMD-TI simulations. The energy of the $\mathrm{A}_1^+$ state is taken as reference. Values in parentheses denote one standard error in the last quoted digits. The last two columns report the ground-state band origins and interchange tunneling splittings: $o_n; i_n$, as defined in Ref.~\citenum{Leforestier-JCP-2012-14305} (a semicolon instead of parentheses is used to avoid confusion with the uncertainty notation).}
\label{tab:water-dimer-splittings}
{\renewcommand{\arraystretch}{1.5}
\begin{ruledtabular}
\begin{tabular*}{\columnwidth}{@{\extracolsep{\fill}}ccccccccc}
  Method  &  $\mathrm{A}_1^+$   &  $\mathrm{B}_1^+$ &  $\mathrm{E}^+$   &  $\mathrm{A}_2^-$   &  $\mathrm{B}_2^-$ &  $\mathrm{E}^-$ & $GS(A'$)-1 & GS(A$'$)-2 \\
\hline
\noalign{\vskip 2pt}
\multicolumn{9}{c}{\textbf{CCpol-8sf}} \\
\hline
PIHMC-EBP &       $0$            &     $0.702(8)$   &   $0.415(19)$     &     $12.74(11)$     &    $13.32(12)$    &    $13.06(15)$  & $0.000; 0.702(8)$  &  $12.68(8); 0.585(12)$\\
 WFN (exact)\cite{Wang-JCP-2018-74108}   &      $0$            &     $0.6853$      &   $0.4086$        &     $12.6306$       &    $13.2117$      &    $12.9866$    & $0.000; 0.685$ & $12.578; 0.581$ \\
\hline
\noalign{\vskip 2pt}
\multicolumn{9}{c}{\textbf{flex-CCpol2025}} \\
\hline
PIHMC-EBP &      $0$            &     $0.765(7)$   &   $0.432(17)$     &     $11.71(10)$     &    $12.35(10)$    &    $12.05(12)$  & $0.000; 0.765(7)$ &   $11.65(7); 0.645(12)$  \\
 WFN (exact)\cite{Jing-JPCL-2025-10923}   &      $0$            &      $0.756$      &    $0.431$        &      $11.609$       &     $12.257$      &     $11.982$    & $0.000; 0.756$ & $11.555; 0.648$ \\
\hline
\noalign{\vskip 2pt}
\multicolumn{9}{c}{\textbf{MB-pol}} \\
\hline
PIHMC-EBP &      $0$            &     $0.796(8)$   &   $0.446(19)$     &     $12.07(10)$     &    $12.74(11)$    &    $12.41(13)$  & $0.000; 0.796(8)$ &  $12.01(7); 0.675(13)$\\
 WFN (approx.)\cite{Babin-JCTC-2013-5395}\footnotemark[1]        &      $0$            &        $0.81$     &        $-$        &       $12.11$       &    $12.80$      &    $-$    & $0.00; 0.81$ & $12.05; 0.69$ \\
 PIMD-TI\cite{Vaillant-JCP-2018-234102}\footnotemark[2]   &      $0$            &      $0.72(4)$      &    $0.37$        &      $9.96$       &     $10.6$      &     $10.2$    & $0.00; 0.72(4)$ & $9.9(12); 0.62(4)$ \\
\end{tabular*}
\end{ruledtabular}
\footnotetext[1]{Variational calculations based on the adiabatic approximation.}
\footnotetext[2]{Values without parentheses have no reported uncertainties in the original literature.}
}
\end{table*}

In constructing the EBP, the presence of 10 conjugacy classes in $G_{16}$ implies, in principle, 10 distinct end states and hence considerable freedom in choosing thermodynamic paths between them.
Symmetry considerations, supported by screening RPI calculations at very small bead numbers, show that several paths are effectively redundant because they probe nearly identical regions of ring-polymer configuration space.
We therefore merge the symmetry operations in selected classes into composite sets, each treated as a single end state in the EBP construction.
The resulting 6 symmetry operation sets, listed in \Tab{tab:sym-sets}, define 6 end states and thus 15 distinct pairwise thermodynamic paths in total.

Panels (a)--(f) of \Fig{fig:water-dimer-kink} show the minimum-action instanton configurations for end states 1--6 in \Tab{tab:sym-sets}, corresponding to identity, acceptor tunneling, geared interchange, bifurcation tunneling, anti-geared interchange, and donor exchange, in the same order.
It is important to note that, unlike density-matrix-element formulations where the endpoints of the RPI configuration are fixed molecular configurations\cite{Richardson-JCP-2011-54109,Richardson-JCP-2011-124109}, symmetrized partition functions specify only the symmetry operation applied at the boundary.
As a result, the instanton solutions are not uniquely tied to a single end state.
For example, while end state 2 is dominated by the minimum-action acceptor-tunneling instanton, a donor-exchange instanton also satisfies the same boundary condition but has a much larger action and is therefore exponentially suppressed.

Although 15 pairwise paths exist in principle, the free-energy calculation does not require all of them.
It suffices to select a network of paths that connects all end states.
To identify an efficient linking network, we first scan all candidate paths using very rough low-$N$ RPI calculations to obtain qualitative estimates of  free-energy profiles, and then choose 5 paths with comparatively small free-energy variation while ensuring full connectivity of the end-state network.
The selected paths are the linear interpolations connecting end-state pairs 1--2, 1--4, 1--5, 4--3, and 5--6, where in each pair the first and second indices denote the initial and final states, respectively, and the initial state has the lower free energy.
Representative instanton configurations at the midpoints $\lambda=0.5$ along these paths are shown in Fig.\,S3 of the Supplementary Material.

In the PIHMC-EBP calculations, we set the temperature to $12\,\mathrm{K}$, as in the previous PIMD study.\cite{Vaillant-JCP-2018-234102}
Based on the numerically exact variational rovibrational levels reported previously on the CCpol-8sf PES, we estimate that vibrational excited-state contributions to the ground-state splitting at this temperature are at most $0.002\,\mathrm{cm^{-1}}$, which is well below the statistical uncertainty of our simulations.
Since the previous PIMD calculations employed a maximum bead number of $N=256$, we begin at this value and then increase $N$ systematically to assess convergence.
For $N=256$, 512, and 768, we use outer time steps $\Delta t=0.3$, 0.2, and $0.2\,\mathrm{fs}$, with propagation lengths $L=100$, 150, and 150, respectively.
Guided by rough RPI estimates of the free-energy variation, the EBP grids for the five selected thermodynamic paths 1--2, 1--4, 1--5, 4--3, and 5--6 are partitioned into 4, 5, 8, 3, and 3 overlapping windows, respectively, giving a total of 23 windows.
The corresponding grid points and bias parameters are reported in the Supplementary Material.
In each long PIHMC run, $5\,\mathrm{ps}$ are discarded for equilibration and $400\,\mathrm{ps}$ are used for data collection.

\Tab{tab:water-dimer-splittings} presents the ground-state energy-level splittings, band origins, and interchange tunneling splittings of the water dimer on the three PESs obtained from the converged PIHMC-EBP calculations (taking the results at $N=768$), together with comparisons to available reference data from other computational approaches.
The complete set of results for all bead numbers and the corresponding convergence tests with respect to $N$ are provided in the Supplementary Material.

As can be seen, the PIHMC-EBP results agree with the numerically exact variational benchmarks for both CCpol-8sf and flex-CCpol2025 within statistical uncertainty, for all six symmetry species and the associated band origins $o_n$ and interchange tunneling splittings $i_n$, where $n=1$ and 2 correspond the $1s$ and $2s$ manifolds, respectively.
This demonstrates the applicability of our method to complex multiwell systems with multiple tunneling pathways.
For MB-pol, the PIHMC-EBP results are also in close agreement with the available results based on the adiabatic approximation.\cite{Babin-JCTC-2013-5395}
The remaining differences are comparable to the present statistical uncertainties, indicating that the error from the adiabatic approximation is of the same order.

The three PESs yield the same qualitative splitting pattern, with $\mathrm{A}_1^+<\mathrm{E}^+<\mathrm{B}_1^+$ in the $1s$ manifold and $\mathrm{A}_2^-<\mathrm{E}^-<\mathrm{B}_2^-$ in the $2s$ manifold, while the individual level splittings and the derived $o_n$ and $i_n$ remain clearly distinguishable from one surface to another.
The present PIHMC-EBP calculations successfully reach a level of precision sufficient to resolve these quantitative differences among the three PESs.

It should be emphasized that only the MB-pol results are obtained directly from the sampled trajectories, whereas the CCpol-8sf and flex-CCpol2025 results are extracted by reweighting the same MB-pol data.
Nevertheless, \Tab{tab:water-dimer-splittings} and the results for all bead numbers in the Supplementary Material show that the statistical uncertainties of the reweighted CCpol-8sf and flex-CCpol2025 values are comparable to those of the directly sampled MB-pol results.
This justifies the reweighting strategy to obtain splittings on multiple PESs from a single set of trajectories without compromising statistical precision, even when the splittings differ substantially among the PESs.

Finally, we compare the present results with the earlier PIMD-TI calculations\cite{Vaillant-JCP-2018-234102} reported for MB-pol.
\Tab{tab:water-dimer-splittings} shows that the PIMD-TI splittings differ markedly from both our PIHMC-EBP results and the available variational values, especially for the band origin $o_2$.
As discussed in \rcite{Vaillant-JCP-2018-234102}, this discrepancy arises primarily from contamination by rotationally excited states in the PIMD-TI calculations.
By contrast, the present formulation employs a rigorous projection onto the rotational ground state and therefore avoids this problem.
Because the PIMD-TI results fail to reproduce the correct tunneling splittings even qualitatively, a comparison of computational efficiency is not meaningful.
It is worth noting that, to our knowledge, the present calculations provide not only the first numerically exact results for the MB-pol surface, but also the first numerically exact determination of the ground-state tunneling splittings of the water dimer within a path-integral framework.

\section{CONCLUDING REMARKS}\label{sec4}
We have proposed the PIHMC-EBP approach for the numerically exact calculation of tunneling splittings in molecular systems.
This method avoids the TI used in earlier PIMD studies and instead enables direct determination of the free-energy profile along the tunneling pathway, from which the splittings are obtained.
Combined with two specialized nonlocal Monte Carlo updates and a reweighting strategy, it provides an efficient and robust route to high-precision tunneling splittings without the cumbersome convergence checks required in PIMD-TI.

The advantages of the new method have been demonstrated by applications to three representative systems.
For malonaldehyde on the PES constructed by Mizukami \textit{et al}.\cite{Mizukami-JCP-2014-144310}, the calculations yield tunneling splittings of $21.31(9)\,\mathrm{cm}^{-1}$ for the normal isotopologue and $2.84(1)\,\mathrm{cm}^{-1}$ for the deuterated species.
For the HCl dimer on a recently constructed high-accuracy PES\cite{Shen-JCP-2025-154307}, the present calculation gives a splitting of $13.1(3)\,\mathrm{cm}^{-1}$.
These results provide the most precise tunneling splittings reported so far for the corresponding PESs.
Compared with previous PIMD-TI studies, the present method reduces the computational cost by several-fold for malonaldehyde and by roughly three orders of magnitude for the HCl dimer.
We have also applied the method to three different water-dimer PESs and obtained numerically exact tunneling splittings within a path-integral framework for the first time.

To suppress severe autocorrelation arising from kink-induced trapping and slow collective inter-bead rotation, we have designed two tailored nonlocal Monte Carlo updates.
These two updates are not restricted to the present framework but can also be incorporated into conventional PIMD simulations.
The kink permutation update may improve the sampling of endpoint configurations in PIMD-TI calculations, while the inter-bead rotation update can enhance the efficiency of calculations involving rotational excitations.
Therefore, these techniques may be useful more broadly in path-integral simulations of molecular spectroscopy.

Another important aspect of the present work is the exploration of a reweighting strategy to obtain results on multiple PESs from a single sampling trajectory.
The water-dimer calculations demonstrate that this strategy can provide statistically reliable tunneling splittings for several PESs simultaneously without loss of statistical precision.
Such a strategy is particularly attractive when systematic comparisons among PESs are desired, for example, in assessing how specific features of a PES affect tunneling observables during PES development.
It also enables the evaluation of  splittings on computationally expensive PESs, for example those without analytic gradients, by sampling trajectories on a more economical reference surface and applying reweighting.
In this way, the cost of rovibrational calculations can become largely independent of the intrinsic cost of evaluating the PES.

Although the systems considered here remain accessible to wavefunction methods, the main strength of path-integral sampling lies in its favorable scaling with system size. Wavefunction-based approaches rapidly become prohibitive as the number of atoms increases, whereas PIHMC and related PIMD techniques can be applied to significantly larger systems without a comparable growth in computational cost. For this reason, path-integral sampling methods, including the approach developed here, are expected to play an increasingly important role in future theoretical studies of tunneling splittings in molecular clusters and larger molecular systems.

\section*{SUPPLEMENTARY MATERIAL}
The Supplementary Material provides water-dimer energy-level splittings on three PESs obtained from PIHMC-EBP calculations with different simulation parameters, and, for all three systems, instanton configurations at the midpoints of the thermodynamic paths ($\lambda=0.5$), plots of fitted action and bias curves together with the corresponding RPI results, and raw grid-point data used to construct the EBPs.

\section*{ACKNOWLEDGEMENTS}
The authors acknowledge financial support from the Swiss National Science Foundation through the project
titled ‘Nonadiabatic effects in chemical reactions’ (Grant Number – 207772). 
The authors thank L\'{e}a Zupan for helpful discussions on the Eckart spring for rotationally excited-state projection.

\section*{AUTHOR DECLARATIONS}
\subsection*{Conflict of Interest}
The authors have no conflicts to disclose.

\subsection*{Author Contributions}
\noindent \textbf{Yu-Chen Wang}: Conceptualization (lead); Methodology (lead); Visualization (lead); Formal analysis (lead); Investigation (lead); Validation (lead); Software (lead); Writing – original draft (lead); Writing – review \& editing (equal).\\
\noindent \textbf{Jeremy O. Richardson}: Conceptualization (supporting); Writing – review \& editing (equal); Funding acquisition (lead).

\section*{Data Availability}
The data that support the findings of this study are available within the article and its supplementary material.

\appendix
\section*{APPENDIX: MBAR for post-processing}\label{app:mbar}

The MBAR method\cite{Shirts-JCP-2008-124105} provides statistically optimal estimators for free-energy differences and ensemble averages by combining samples from multiple, possibly overlapping ensembles.
Since its introduction, it has become a standard tool for free-energy calculations in molecular simulation, particularly in biomolecular applications where data are routinely collected across many overlapping thermodynamic states.
In the present work, MBAR provides a natural bridge between the windowed PIHMC sampling described in \Sec{sec:windowing} and \Sec{sec:pihmc} and the final free-energy and prefactor estimates needed for symmetrized partition functions.

Here, we summarize the minimal theory required for our implementation.
Consider a set of source ensembles (windows) indexed by $w=1,\dots,W$, each corresponds to one of the overlapping windows defined in \Sec{sec:windowing}.
The $w$-th window is associated with the local EBP $U^{(w)}(\bm{q})$.
Assume that we have collected $N_w$ ring-polymer configurations in window $w$, and denote the total number of configurations as $N_\mathrm{tot}=\sum_{w}N_w$.
For convenience, all configurations are labeled collectively as $\{\bm{q}_{r}\}$, where $\bm{q}_{r}$ denotes the $r$-th configuration, regardless of the window from which it originates.
Each source ensemble is characterized by a reduced potential,
\begin{equation}
u^{(w)}(\bm{q})=\beta_N U^{(w)}(\bm{q}),
\end{equation}
where the notation $u^{(w)}(\bm{q})$ should not be confused with the prefactor $u_{P}^{(J)}(\bm{q})$ defined in \Sec{sec:zpj}.
For any thermal ensembles of interest, we use the index $w$ for sampled source windows and $s$ for an arbitrary state (which may coincide with a source window); $t$ denotes a target state of interest.
For a state $s$ with the reduced potential $u^{(s)}(\bm{q})$, the dimensionless free energies\footnote{In MBAR, the set of $f^{(s)}$ is determined only up to a common additive constant; only free-energy differences are physically meaningful.} are formally defined by $f^{(s)}=-\ln Z^{(s)}$ with $Z^{(s)}=\int \mathrm{d}\bm{q}\, e^{-u^{(s)}(\bm{q})}$.
Given the pooled samples $\{\bm{q}_{r}\}$ from all source windows, MBAR yields the following estimator for $f^{(s)}$:
\begin{widetext}
\begin{equation}
\label{eq:mbar-sc}
f^{(s)}=-\ln
\sum_{r=1}^{N_\mathrm{tot}}
\frac{\exp\left[-u^{(s)}(\bm{q}_{r})\right]}
{\sum_{w'=1}^{W} N_{w'}\exp\left[f^{(w')}-u^{(w')}(\bm{q}_{r})\right]}.
\end{equation}
\end{widetext}
When $s$ is chosen to be one of the source windows, $s=w$, this relation yields a closed set of self-consistent equations for the unknown free energies $\{f^{(w)}\}_{w=1}^{W}$, which can be obtained by solving these equations iteratively or using numerical optimization methods.
Once $\{f^{(w)}\}$ are determined, the free energy of any target state $t$ can be obtained by the same expression with $s=t$, provided that $u^{(t)}$ can be evaluated on all pooled samples.
Free-energy differences between any two states then follow as $\Delta f_{ss'}=f^{(s')}-f^{(s)}$.

Ensemble averages of an observable $O(\bm{q})$ under a target distribution are given by 
\begin{equation}\label{eq:mbar-obs}
\langle O\rangle_t = \sum_{r=1}^{N_\mathrm{tot}}w^{(t)}(\bm{q}_{r})O(\bm{q}_{r}),
\end{equation}
where
\begin{equation}
w^{(t)}(\bm{q}_{r})=\frac{\exp\left[{f^{(t)}-u^{(t)}(\bm{q}_{r})}\right]}
{\sum_{w'=1}^{W}N_{w'}\exp\left[f^{(w')}-u^{(w')}(\bm{q}_{r})\right]}
\end{equation}
is the normalized weight.
In practice, applying MBAR therefore requires storing the reduced potentials $u^{(w)}(\bm{q}_{r})$ for all source windows and $u^{(t)}(\bm{q}_{r})$ for any desired target, as well as the values of observables $O(\bm{q}_{r})$ to be averaged.

In our tunneling splitting calculations, the target reduced potentials $u^{(t)}(\bm{q})$ are defined according to the ring-polymer potentials $\widetilde{U}_{P}(\bm{q})$ [\Eq{eq:up-tilde}] associated with the symmetrized partition functions of all symmetry operations $\hat{P}\in\mathcal{G}$.
In \Eq{eq:mbar-obs}, the observable $O(\bm{q})$ can be any quantity of interest, and in the present application it is most often taken to be the prefactor $u_{P}^{(J)}(\bm{q})$, whose ensemble averages enter the estimators of the $J$-projected symmetrized partition functions.

When the number of samples or windows is large, solving the full set of self-consistent equations can be slow.
We therefore initialize the global MBAR solver by first applying MBAR to small subsets of two to several neighboring windows to obtain local free-energy differences, and then combining these differences by a least-squares fit to produce a globally consistent initial guess.
This initialization substantially accelerates convergence.

According to the above description, one might expect that one has to decide in advance which target potentials will be analyzed, so that the corresponding reduced potentials can be evaluated during the PIHMC run.
This would indeed be the case if evaluating the reduced potentials required the full ring-polymer configuration.
In the present setting, however, all source and target ring-polymer potentials share the same open-chain term $U_{N}^{\mathrm{(open)}}(\bm{q})$ and differ only in the boundary spring term, which depends solely on the endpoint beads $\mathbf{q}^{(1)}$ and $\mathbf{q}^{(N)}$.
As a consequence, we may shift all reduced potentials by the same configuration-dependent term,
\begin{equation}
\bar{u}^{(s)}(\bm{q}) = u^{(s)}(\bm{q}) - \beta_NU^\mathrm{(open)}(\bm{q}),
\end{equation}
which leaves the MBAR equations and the resulting free-energy differences unchanged, because the common factor cancels in the numerator and denominator for each sampled configuration in \Eq{eq:mbar-sc}.
The prefactor $u_{P}^{(J)}(\bm{q})$ is likewise determined by the endpoint configurations.
We therefore evaluate MBAR using only the boundary contributions, and avoid the need to recompute $U^\mathrm{(open)}(\bm{q})$ during post-processing.

Concretely speaking, we store $\mathbf{q}^{(1)}$ and $\mathbf{q}^{(N)}$  for every sampled configuration in each window.
After completing all PIHMC runs, a post-processing program reads these endpoint data and constructs the source reduced potentials for the window ensembles from the corresponding boundary terms, omitting the common open-chain contribution $U^\mathrm{(open)}(\bm{q})$ which cancels exactly in the MBAR ratios.
In the same post-processing step, the program evaluates the target reduced potentials for all symmetry operations $\hat{P}\in\mathcal{G}$, together with the associated prefactors needed for the symmetrized partition functions.
Solving the MBAR self-consistent equations then yields the statistically optimal estimates of the free energies of all target distributions and provides the required prefactor averages.
These results determine the ratios of symmetrized partition functions and, upon substitution into \Eq{eq:delta-ZPJ}, the tunneling splittings.

This workflow separates the calculation into two independent stages, PIHMC sampling and MBAR analysis.
Once the PIHMC sampling has been completed and the endpoint configurations have been stored, additional quantities can be obtained by repeating only the post-processing, for example by evaluating prefactors for additional $J$ values, without rerunning PIHMC trajectories.

If the PIHMC sampling is combined with the PES reweighting strategy described in \Sec{sec:reweight-pes} to obtain results on multiple PESs from a single run on the reference PES, $V_{\mathrm{ref}}$, the MBAR post-processing must account explicitly for the PES dependence of the physical potential term.
Accordingly, for each sampled configuration $\bm{q}$ and for each PES, $V_\alpha$, of interest, we evaluate and store on-the-fly the bead-summed physical potentials
\begin{equation}
V_{\alpha}^{\mathrm{RP}}(\bm{q})=\sum_{i=1}^{N} V_\alpha\!\left(\mathbf{q}^{(i)}\right).
\end{equation}
Then, for any state $s$ analyzed on PES $V_\alpha$, the reduced potential used in MBAR is
\begin{equation}
\bar{u}^{(s,\alpha)}(\bm{q}) =
\bar{u}^{(s)}(\bm{q}) 
+\beta_N\left[V_{\alpha}^{\mathrm{RP}}(\bm{q})-V_{\mathrm{ref}}^{\mathrm{RP}}(\bm{q})\right],
\end{equation}
where $V_{\mathrm{ref}}^{\mathrm{RP}}(\bm{q})$ is the bead-summed reference physical potential.
This construction ensures that, when evaluating a target reduced potential on $V_\alpha$, the physical-potential contribution is correctly included even though the PIHMC dynamics was carried out on $V_{\mathrm{ref}}$.

Finally, statistical uncertainties of all MBAR-derived quantities, including the final tunneling-splitting values, are estimated using circular-block bootstrapping to account for correlations in the sampled time series.
To avoid repeated calculations of the reduced potentials and prefactors required by MBAR for each bootstrap sample, these quantities are first evaluated for all sampled configurations and stored in the original sampling order, and the bootstrap resampling is then performed on these precomputed quantities rather than on the configurations themselves.
Further details of the bootstrap procedure are given in \Sec{sec:error}.

\bibliography{wyc-pihmc,references}

\clearpage

\end{document}


\begin{CJK*}{UTF8}{gbsn}

\title{Supplementary material: Exact tunneling splittings from path-integral hybrid Monte Carlo with enveloping bridging potentials}

\author{Yu-Chen Wang (汪宇晨)}
\email{wangyuc@phys.chem.ethz.ch}

\author{Jeremy O. Richardson}
\affiliation{\mbox{Institute of Molecular Physical Science, ETH Z\"{u}rich, 8093 Z\"{u}rich, Switzerland}}

\maketitle
\end{CJK*}

\noindent This Supplementary Material provides water-dimer energy-level splittings on three PESs obtained from PIHMC-EBP calculations with different simulation parameters and, for all three systems studied in this work, instanton configurations at the midpoints of the thermodynamic paths ($\lambda=0.5$) and plots of fitted action and bias curves together with the corresponding RPI results. The raw grid-point data used to construct the EBPs for all three systems are provided in Supplementary File \texttt{ebp\_grid.zip}.

\begin{table}[H]
\caption{Energy-level splittings (cm$^{-1}$) of the water dimer on the \textbf{CCpol-8sf}\cite{Leforestier-JCP-2012-14305,Wang-JCP-2018-74108} PES from PIHMC-EBP simulations at $T=12\,\mathrm{K}$ with different bead numbers $N$. The energy of the $\mathrm{A}_1^+$ state is taken as reference. Values in parentheses denote one standard error in the last quoted digits. In the bottom part of the table, the converged PIHMC-EBP results (taken from $N=768$, highlighted in bold in the table) are compared with numerically exact variational (WFN) results on the same PES. The last two columns report the ground-state band origins and interchange tunneling splittings: $o_n; i_n$, as defined in Ref.~\citenum{Leforestier-JCP-2012-14305} (a semicolon instead of parentheses is used to avoid confusion with the uncertainty notation).}
\label{tab:MB-pol}
{\renewcommand{\arraystretch}{1.5}
\begin{ruledtabular}
\begin{tabular*}{\columnwidth}{@{\extracolsep{\fill}}ccccccccc}
  $N$  &  $\mathrm{A}_1^+$   &  $\mathrm{B}_1^+$ &  $\mathrm{E}^+$   &  $\mathrm{A}_2^-$   &  $\mathrm{B}_2^-$ &  $\mathrm{E}^-$ & $GS(A'$)-1 & GS(A$'$)-2 \\
\hline
  256  &      $0$            &     $0.742(6)$   &   $0.396(13)$     &     $13.34(10)$     &    $13.95(11)$    &    $13.86(13)$  & $0.000; 0.742(6)$ &   $13.25(7); 0.640(12)$ \\
  512  &      $0$            &     $0.694(6)$   &   $0.397(18)$     &     $12.88(11)$     &    $13.46(12)$    &    $13.33(14)$  & $0.000; 0.694(6)$ & $12.82(8); 0.583(12)$ \\
  768  &      $0$            &     $\B{0.702(8)}$   &   $\B{0.415(19)}$     &     $\B{12.74(11)}$     &    $\B{13.32(12)}$    &    $\B{13.06(15)}$  &  $\B{0.000; 0.702(8)}$ &  $\B{12.68(8); 0.585(12)}$ \\
\hline
Final &       $0$            &     $0.702(8)$   &   $0.415(19)$     &     $12.74(11)$     &    $13.32(12)$    &    $13.06(15)$  & $0.000; 0.702(8)$  &  $12.68(8); 0.585(12)$\\
\hline
 WFN\cite{Wang-JCP-2018-74108}   &      $0$            &     $0.6853$      &   $0.4086$        &     $12.6306$       &    $13.2117$      &    $12.9866$    & $0.000; 0.685$ & $12.578; 0.581$ \\
\end{tabular*}
\end{ruledtabular}
}
\end{table}

\begin{table}[H]
\caption{Energy-level splittings (cm$^{-1}$) of the water dimer on the \textbf{flex-CCpol2025}\cite{Jing-JPCL-2025-10923} PES from PIHMC-EBP simulations at $T=12\,\mathrm{K}$ with different bead numbers $N$. The energy of the $\mathrm{A}_1^+$ state is taken as reference. Values in parentheses denote one standard error in the last quoted digits. In the bottom part of the table, the converged PIHMC-EBP results (taken from $N=768$, highlighted in bold in the table) are compared with numerically exact variational (WFN) results on the same PES. The last two columns report the ground-state band origins and interchange tunneling splittings: $o_n; i_n$, as defined in Ref.~\citenum{Leforestier-JCP-2012-14305} (a semicolon instead of parentheses is used to avoid confusion with the uncertainty notation).}
\label{tab:MB-pol}
{\renewcommand{\arraystretch}{1.5}
\begin{ruledtabular}
\begin{tabular*}{\columnwidth}{@{\extracolsep{\fill}}ccccccccc}
  $N$  &  $\mathrm{A}_1^+$   &  $\mathrm{B}_1^+$ &  $\mathrm{E}^+$   &  $\mathrm{A}_2^-$   &  $\mathrm{B}_2^-$ &  $\mathrm{E}^-$ & $GS(A'$)-1 & GS(A$'$)-2 \\
\hline
  256  &      $0$            &     $0.778(7)$   &   $0.409(13)$     &     $12.40(9)$     &    $13.08(10)$    &    $12.95(11)$  &  $0.000; 0.778(7)$ & $12.35(7); 0.681(13)$\\
  512  &      $0$            &     $0.751(7)$   &   $0.413(15)$     &     $11.88(9)$     &    $12.51(10)$    &    $12.31(11)$  & $0.000; 0.751(7)$ &   $11.82(7); 0.638(12)$ \\
 768   &      $0$            &     $\B{0.765(7)}$   &   $\B{0.432(17)}$     &     $\B{11.71(10)}$     &    $\B{12.35(10)}$    &    $\B{12.05(12)}$  &  $\B{0.000; 0.765(7)}$ &  $\B{11.65(7); 0.645(12)}$ \\
\hline
Final &      $0$            &     $0.765(7)$   &   $0.432(17)$     &     $11.71(10)$     &    $12.35(10)$    &    $12.05(12)$  & $0.000; 0.765(7)$ &   $11.65(7); 0.645(12)$  \\
\hline
 WFN\cite{Jing-JPCL-2025-10923}   &      $0$            &      $0.756$      &    $0.431$        &      $11.609$       &     $12.257$      &     $11.982$    & $0.000; 0.756$ & $11.555; 0.648$ \\
\end{tabular*}
\end{ruledtabular}
}
\end{table}

\begin{table}[H]
\caption{Energy-level splittings (cm$^{-1}$) of the water dimer on the \textbf{MB-pol}\cite{Babin-JCTC-2013-5395} PES from PIHMC-EBP simulations at $T=12\,\mathrm{K}$ with different bead numbers $N$. The energy of the $\mathrm{A}_1^+$ state is taken as reference. Values in parentheses denote one standard error in the last quoted digits. In the bottom part of the table, the converged PIHMC-EBP results (taken from $N=768$, highlighted in bold in the table) are compared with \textbf{approximate} variational (WFN) results and the PIMD-TI results on the same PES. The last two columns report the ground-state band origins and interchange tunneling splittings: $o_n; i_n$, as defined in Ref.~\citenum{Leforestier-JCP-2012-14305} (a semicolon instead of parentheses is used to avoid confusion with the uncertainty notation).}
\label{tab:MB-pol}
{\renewcommand{\arraystretch}{1.5}
\begin{ruledtabular}
\begin{tabular*}{\columnwidth}{@{\extracolsep{\fill}}ccccccccc}
  $N$  &  $\mathrm{A}_1^+$   &  $\mathrm{B}_1^+$ &  $\mathrm{E}^+$   &  $\mathrm{A}_2^-$   &  $\mathrm{B}_2^-$ &  $\mathrm{E}^-$ & $GS(A'$)-1 & GS(A$'$)-2 \\
\hline
  256  &      $0$            &     $0.820(7)$   &   $0.426(12)$     &     $12.86(10)$     &    $13.58(11)$    &    $13.44(12)$  & $0.000; 0.820(7)$ &   $12.81(7); 0.722(14)$ \\
  512  &      $0$            &     $0.783(7)$   &   $0.424(15)$     &     $12.27(10)$     &    $12.94(10)$    &    $12.73(12)$  & $0.000; 0.783(7)$ &  $12.22(7); 0.671(13)$ \\
 768   &      $0$            &     $\B{0.796(8)}$   &   $\B{0.446(19)}$     &     $\B{12.07(10)}$     &    $\B{12.74(11)}$    &    $\B{12.41(13)}$  & $\B{0.000; 0.796(8)}$ &  $\B{12.01(7); 0.675(13)}$ \\
\hline
Final &      $0$            &     $0.796(8)$   &   $0.446(19)$     &     $12.07(10)$     &    $12.74(11)$    &    $12.41(13)$  & $0.000; 0.796(8)$ &  $12.01(7); 0.675(13)$\\
\hline
 WFN (approx.)\cite{Babin-JCTC-2013-5395}        &      $0$            &        $0.81$     &        $-$        &       $12.11$       &    $12.80$      &    $-$    & $0.00; 0.81$ & $12.05; 0.69$ \\
\hline
 PIMD-TI\cite{Vaillant-JCP-2018-234102}   &      $0$            &      $0.72(4)$      &    $0.37$        &      $9.96$       &     $10.6$      &     $10.2$    & $0.00; 0.72(4)$ & $9.9(12); 0.62(4)$ \\
\end{tabular*}
\end{ruledtabular}
}
\end{table}

\newpage

\begin{figure*}[htbp]
    \centering
    \includegraphics[width=0.5\columnwidth]{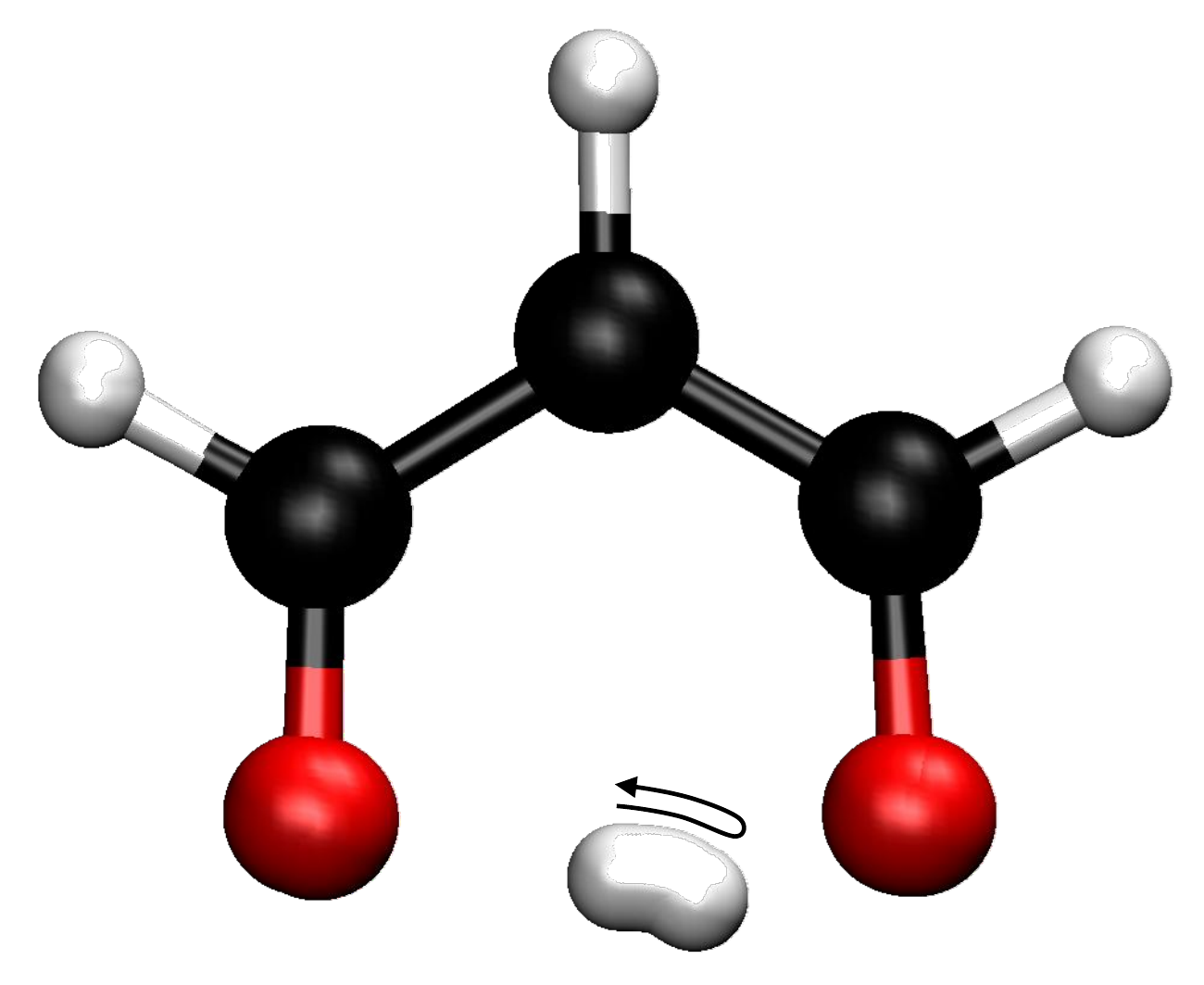}
    \caption{Instanton configuration for the malonaldehyde at the midpoint of the thermodynamic path $\lambda=0.5$. The arrow indicates the direction of proton transfer.}
    \label{fig:mhth-middle}
\end{figure*}

\begin{figure*}[htbp]
    \centering
    \includegraphics[width=0.5\columnwidth]{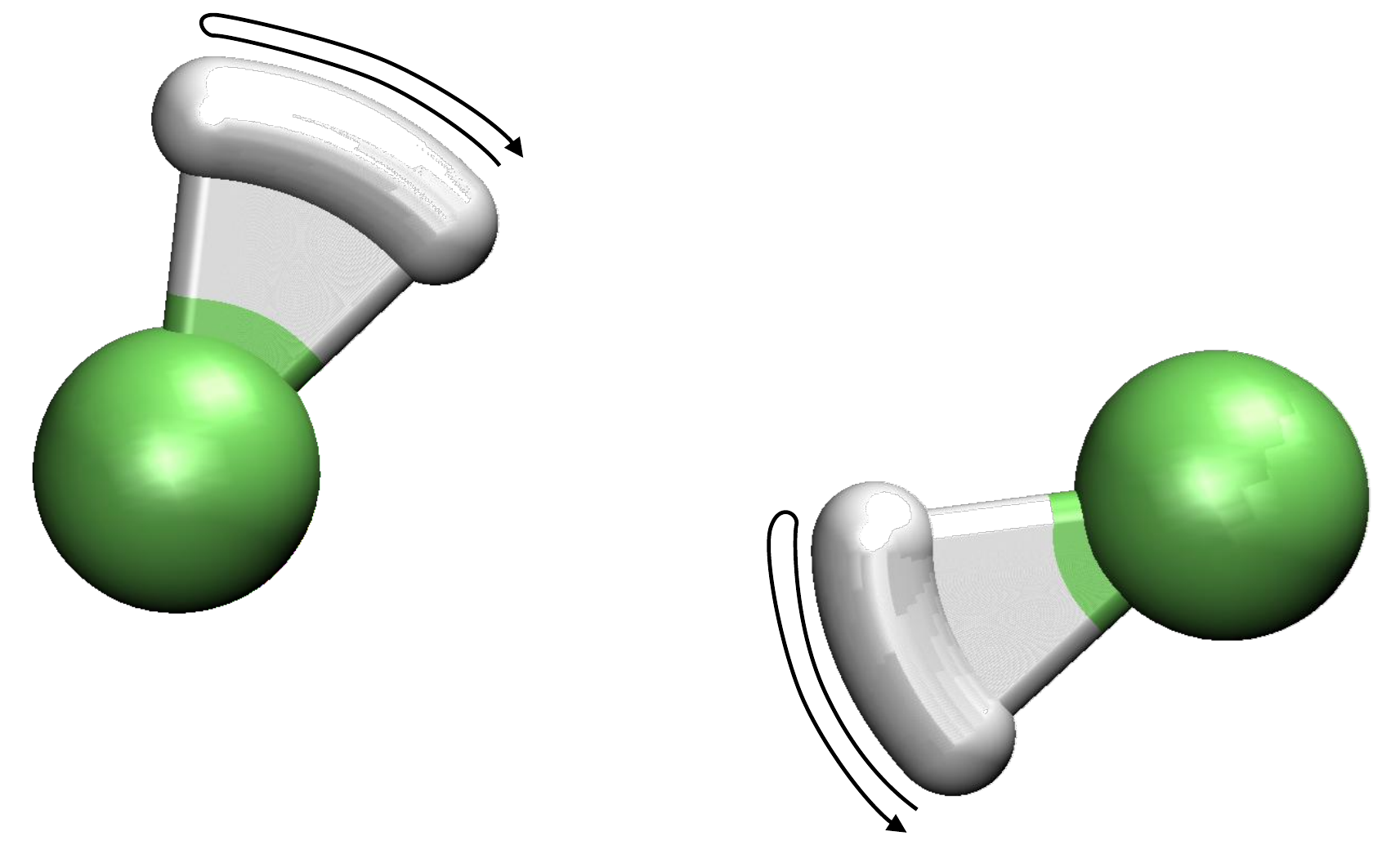}
    \caption{Instanton configuration for the HCl dimer at the midpoint of the thermodynamic path $\lambda=0.5$. The arrow indicates the direction of motion.}
    \label{fig:hcl2-middle}
\end{figure*}

\begin{figure*}[htbp]
    \centering
    \includegraphics[width=0.98\linewidth]{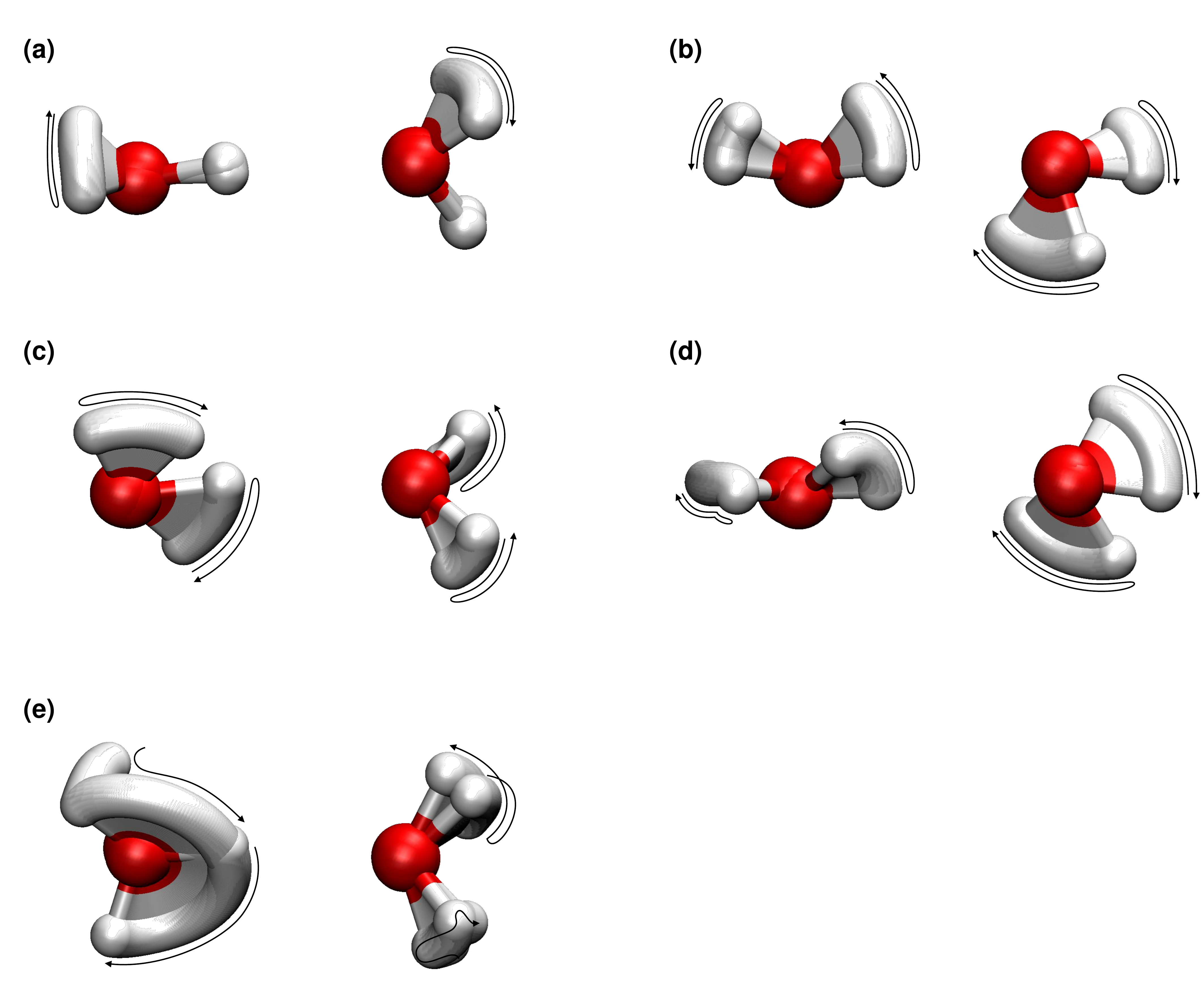}
    \caption{Minimum-action instanton configurations for the water dimer at the midpoint $\lambda=0.5$ of the five thermodynamic paths selected for the EBP construction. Each panel corresponds to the linear interpolation between the composite operation sets listed in Tab.\,VI in the main text: (a) sets 1--2, identity to acceptor tunneling, (b) sets 1--4, identity to bifurcation tunneling, (c) sets 1--5, identity to anti-geared interchange, (d) sets 4--3, bifurcation to geared tunneling, (e) sets 5--6, anti-geared interchange to donor exchange.}
    \label{fig:water-dimer-middle}
\end{figure*}

\clearpage

\begin{figure*}[htbp]
    \centering
    \includegraphics[width=0.5\columnwidth]{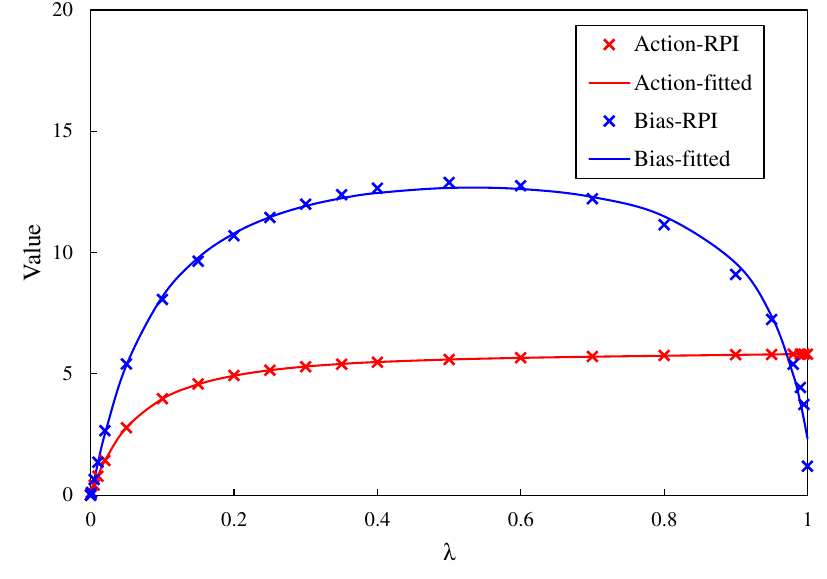}
    \caption{RPI-calculated Euclidean action values, $\beta_NU_\mathrm{m}(\tilde{\bm{q}};\lambda)$, and dimensionless biases, $\tilde{\epsilon}_l$ [Eq.~(29)], for malonaldehyde at the preliminary $\lambda$ grid points, together with the corresponding double-hill-model fits [Eq.~(28)].}
    \label{fig:fitted-mht-h}
\end{figure*}

\begin{figure*}[htbp]
    \centering
    \includegraphics[width=0.5\columnwidth]{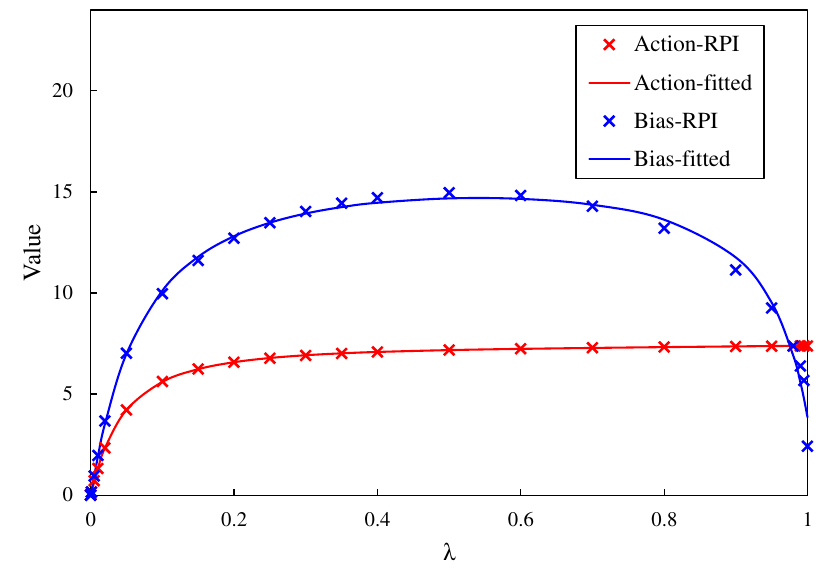}
    \caption{RPI-calculated Euclidean action values, $\beta_NU_\mathrm{m}(\tilde{\bm{q}};\lambda)$, and dimensionless biases, $\tilde{\epsilon}_l$ [Eq.~(29)], for deuterated malonaldehyde at the preliminary $\lambda$ grid points, together with the corresponding double-hill-model fits [Eq.~(28)].}
    \label{fig:fitted-mht-d}
\end{figure*}

\begin{figure*}[htbp]
    \centering
    \includegraphics[width=0.5\columnwidth]{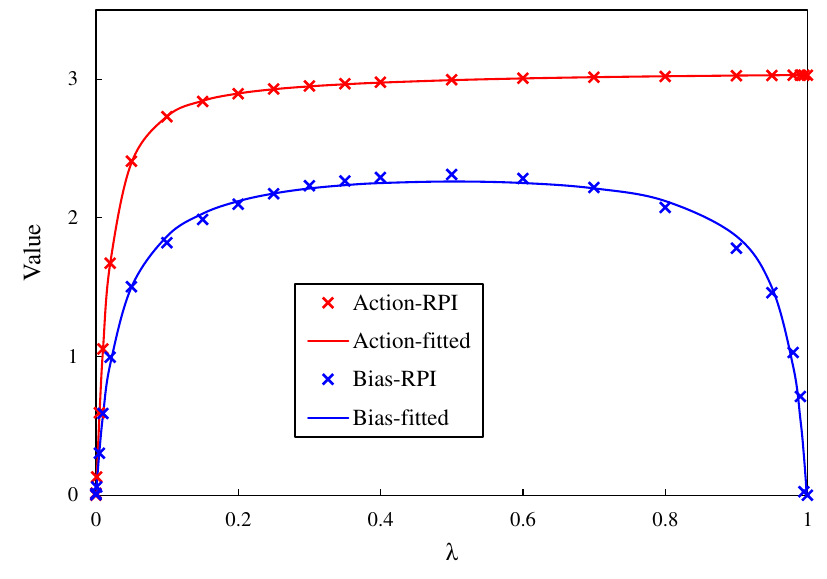}
    \caption{RPI-calculated Euclidean action values, $\beta_NU_\mathrm{m}(\tilde{\bm{q}};\lambda)$, and dimensionless biases, $\tilde{\epsilon}_l$ [Eq.~(29)], for the HCl dimer at the preliminary $\lambda$ grid points, together with the corresponding double-hill-model fits [Eq.~(28)].}
    \label{fig:fitted-hcl-dimer}
\end{figure*}

\begin{figure*}[htbp]
    \centering
    \includegraphics[width=0.98\columnwidth]{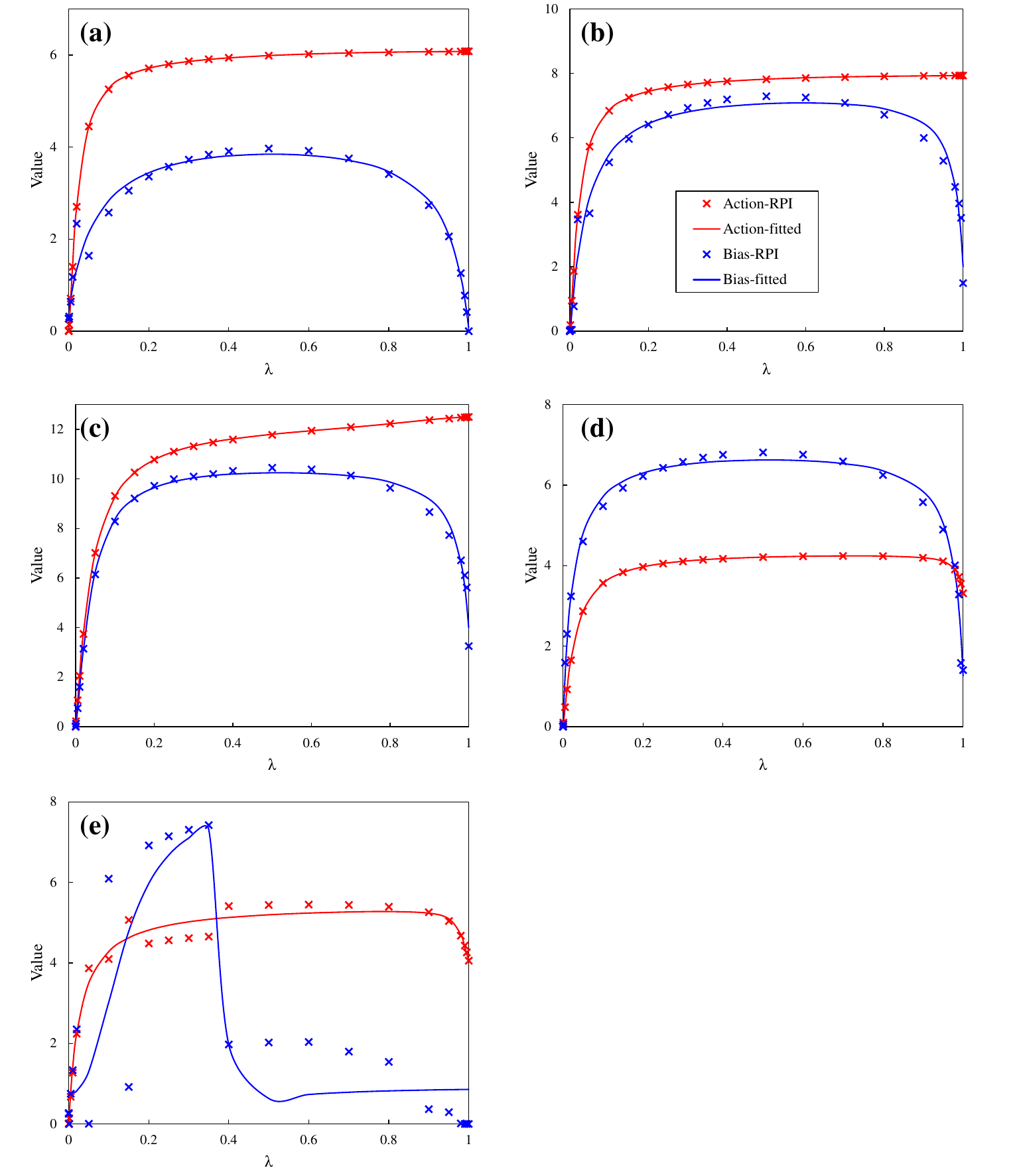}
    \caption{RPI-calculated Euclidean action values, $\beta_NU_\mathrm{m}(\tilde{\bm{q}};\lambda)$, and dimensionless biases, $\tilde{\epsilon}_l$ [Eq.~(29)], for the water dimer at the preliminary $\lambda$ grid points along the selected thermodynamic paths, together with the corresponding double-hill-model fits [Eq.~(28)]. Panels (a)–(e) correspond to the end-state pairs 1--2, 1--4, 1--5, 4--3, and 5--6, respectively. It is noted that the RPI results in panel (e) exhibit discontinuities, particularly in the bias values, which arise because the optimizations at some grid points converge to local rather than global minima. The fitted action curve nevertheless captures the overall trend and provides a reasonable basis for selecting the EBP grid points. Since the bias values estimated by RPI serve only as initial guesses and are subsequently refined by short PIHMC simulations, the sampling performance and convergence are not noticeably affected, demonstrating the robustness of the present approach.}
    \label{fig:fitted-water-dimer}
\end{figure*}

\clearpage

\section*{REFERENCES}

\bibliography{wyc-pihmc,references}